\documentclass[twocolumn,preprintnumbers,amsmath,amssymb,pra]{revtex4}

\usepackage{graphicx,amsmath}
\usepackage{dcolumn}
\usepackage{bm}
\usepackage{amssymb}
\usepackage{epstopdf}
\usepackage{color}

\begin{document}

\title{Density matrix and space-time distributions of the electronic density and current at fast pulsed photoemission through a double quantum well}

\begin{abstract}
Within the framework of the density matrix method, general formulas obtained that are convenient for describing fast pulsed photoemission that occurs in a time less than or on the order of the times of relaxation processes inside the photocathode. Expressions for the elements of the density matrix are found by solving the kinetic equation that takes into account the alternating electromagnetic field of light pumping and inelastic scattering of electrons. The derived formulas are applied for the numerical-analytical study of a one-dimensional model of wave-like spatiotemporal modulation of a photoelectron pulse of suitable duration during its passage through a double-well quantum-well heterostructure deposited on a volumetric planar photocathode. This modulation is a quantum beat that occurs as a result of excitation and subsequent slow oscillatory decay of the superposition of the doublet of quasi-stationary states of the heterostructure. It is possible to provide prolongation of generation and even amplification of waves of charge density and current density of photoelectrons when the photocathode is exposed to a periodic sequence of light pulses.
\end{abstract}


\date{\today }

\author{Yu. G. Peisakhovich}
\affiliation{Novosibirsk State Technical University, Novosibirsk,
Russia}
\author{A. A. Shtygashev} \email{shtygashev@corp.nstu.ru}
\affiliation{Novosibirsk State
Technical University, Novosibirsk, Russia}


 \maketitle

\section{Introduction}\label{intr}
The creation of laser light sources capable of generating ultrashort pulses of picosecond, femtosecond and even attosecond duration has led to the intensive development of spectroscopy and high technologies in the corresponding frequency ranges \cite{Rost2011}-\cite{Dabr2017}. Most often, the purpose of using this high-frequency technique is to obtain spectroscopic information on rapidly proceeding processes in rarefied and condensed media: on the dynamics of the motion of electrons in atoms and molecules, in metallic and semiconductor solids, on the processes of photoexcitation and relaxation of various vibrations in these systems, on the kinetics chemical reactions, etc. This includes, in particular, pulsed photoemission techniques such as two-photon time-resolved photoemission spectroscopy \cite{Soli2003}-\cite{Stol2004}, quantum beat spectroscopy \cite{Garr1998}-\cite{Silk2015} and other methods of linear and nonlinear photoemission probing of matter \cite{Reut2019},\cite{Ferr2009}. Pulsed photoemission techniques are also used to obtain the maximum quantum yield of semiconductor photocathodes when creating highly efficient electron photoinjectors and photomultipliers \cite{Herr1996}-\cite{Gerc2012}.

The analysis of the corresponding experimental data at present \cite{Huff2010},\cite{Soli2003} is carried out mainly on the basis of the achievements of the theory of stationary photoemission, which was intensively developed in the late 50s - early 70s of the last century. The most widely used are the semiphenomenological three-step model of Spicer's photoemission \cite{Spic1958}, \cite{Spic1993} (including different versions of the application of the Fermi golden rule for estimating the probabilities of light absorption \cite{Huff2010}, \cite{Soli2003}]) and the formally more rigorous, but much more difficult to interpret microscopic theory based on the application of the diagram technique for nonequilibrium Green's functions (one-step model) \cite{Caro1973}-\cite{Brau2016}. The microscopic theory of stationary photoemission from crystals includes the calculation of the photocurrent in the second order of the perturbation theory in the electromagnetic field. In this case, the field is usually considered monochromatic with a certain frequency  $\omega$, and it is in the second order that a constant component appears in the current (the goal of calculations and the most frequent experimental measurements), therefore, complete averaging over time is performed from the very beginning. The magnitude of this averaged stationary current and its frequency-energy distribution are determined, first of all, by the energy spectrum of electrons, as well as by the processes of elastic and inelastic scattering in the near-surface region of the photocathode. If the mean free path of photoelectrons is small compared to the depth of photoexcitation, then taking inelastic processes into account becomes especially important, although it is described by difficult-to-estimate higher-order scattering diagrams, which can sometimes be estimated by series and sums expressed in terms of phenomenological lengths and times of electron free path \cite{Caro1973}. From the phenomenological considerations of the three-step model, it follows that at excitation energies of the order of several electron-volts, the main mechanisms of photoelectron scattering and the characteristics of the photocurrent in metals and semiconductors are very different. In bulk metal photocathodes, where electron-electron scattering predominates the response time of the photocurrent to photoexcitation and the relaxation time after switching off the illumination $\tau  \sim 10^{ - 15}  - 10^{ - 14} $ s is much shorter than in semiconductor photocathodes, where the main thing is electron-phonon scattering and $\tau  \sim 10^{ - 13}  - 10^{ - 12} $
 s; it is especially much less than in photocathodes with negative electron affinity, where a large photo yield is determined by slow processes of thermalization accumulation of photoexcited electrons at the bottom of the conduction band and their diffusion to the surface, which leads to $\tau  \sim 10^{ - 10}  - 10^{ - 9} $ s \cite{Hart1999}, \cite{Aule2002}, \cite{Spic1993}. In the presence of quantum-size films, superlattices, surface levels of the image potential on the surface of metals and semiconductors, some peaks and thresholds are observed in the photoemission energy and angular distributions, indicating the formation of resonant quasi-stationary states with energies below and above the vacuum level \cite{Houd1985}-\cite{Chia2000}. 
In \cite{Gerc2012}, it was demonstrated that the use of strained semiconductor superlattices as elements of photoemitter with negative electron affinity leads to such a rearrangement of the spectrum and a change in the dynamics of electrons in the active region, which increase the quantum yield and the degree of polarization of photoelectrons, significantly changing the relaxation times.
 
For a productive theoretical description of nonstationary pulsed photoemission, three main approaches are used: 1) to calculate the time-dependent probabilities of fast femtosecond pumping-probing processes in two-photon photoemission with time resolution, the  technique of the Keldysh's nonequilibrium Green's functions is used \cite{Brau1996}, \cite{Brau2016} in line with the development of the one-step model \cite{Pend1976}, \cite{Hopk1985}; 2) to interpret the same probabilities, as well as to describe quantum beats in such systems, the density matrix method with the solution of the Bloch equations for the two-level \cite{Hert1996} and three-level \cite{Baue1999}, \cite{Klam2001} models is currently most often used; 3)  to describe slower picosecond and nanosecond relaxation processes in semiconductor photocathodes with negative electron affinity, the diffusion equation  is solved \cite{Hart1999}, \cite{Aule2002} within the framework of a three-step model, and the density matrix method is sometimes used \cite{Gerc2012}.

It should be noted that due to the complexity of both the physical processes themselves and their mathematical description, in all these approaches, at one stage or another, some strictly unprovable simplifications based on various physical assumptions are introduced. As a result of such simplifications, it is possible to implement model calculations that provide a relatively satisfactory qualitative and semi-quantitative explanation of the corresponding experimental data. 

In addition to pulsed spectroscopic sensing of matter, of interest is the problem of generating high-frequency oscillations and waves of electron density and current by converting in them ultrashort laser excitation pulses acting on the system. Thin-film nanoscale heterostructures in the form of a double quantum well with tunnel-transparent walls for electrons are suitable for this; such a system has doublets of relatively close stationary or resonance quasi-stationary levels in the energy spectrum of the transverse motion of electrons. Coupled oscillations of mixed doublet resonance states can manifest themselves as quantum beats of the space-time distributions of the probability density and the current density of electrons, the energies of which belong to a narrow band that includes the doublet. Such beats usually accompany a quantum transient process \cite{Camp2009}-\cite{Romo2002} after a single pulse excitation and last for the lifetime of quasi-stationary states, which can be much longer than the time period of these beats if the transparency of the barriers is sufficiently low. In the previous article \cite{Peis2021}, we investigated the case when the population of the doublet was provided by scattering of an electron wave packet incident on the system from the outside. Such a problem is rigorously formulated and solved numerically-analytically in terms of pure quantum-mechanical states of the scattering problem, making it possible to estimate the contributions of the main features and to understand many details of the process that are important in more complex cases. Of interest is also the question of the photoexcitation of quasi-stationary states of electrons in potential wells of such a heterostructure using a short photoemission pulse.

In this article, our first goal is to obtain some general formulas that are convenient for describing fast pulsed photoemission that occurs in a time less than or on the order of the times of relaxation processes inside the photocathode. For this, it is advisable to apply a variant of the density matrix method, which was developed to describe dynamic processes in metals and semiconductors \cite{Ross2002}, \cite{Kope1970}-\cite{Nabu1976}. In the apparatus of the density matrix, mixed states are operated taking into account the influence of an external high-frequency electromagnetic pumping field and the interaction of electrons with surrounding particles.

Breaking off the chain of equations for the density matrix in the second order in the light electric field, one can obtain approximate expressions for the space-time distributions of the electronic probability and current densities for weak inelastic incoherent processes, which correspond to the approximate formulas of the perturbation theory for the steady-state photoemission current \cite{Caro1973}. On the basis of this method, earlier in the joint work of V.M. Nabutovsky and one of the authors \cite{Nabu1976}, a theory of threshold features of the frequency-energy distributions of photoelectrons was developed, in this case, only the stationary photocurrent was calculated within the framework of the three-step model, and only the time-independent diagonal elements of the density matrix determined in the second order of the perturbation theory in the electric field were taken into account. In the nonstationary case, it is required to calculate both the diagonal and off-diagonal elements of the density matrix, which depend on time in accordance with the quantum kinetic equation describing the effect of the alternating electromagnetic field of the pump pulse, as well as various inelastic processes partially responsible for relaxation. 

The products of the density matrix elements and the coordinate-dependent elements of the probability density or current "matrices" summed over the states of the registered dedicated energy band give measurable pulsed distributions of the electronic densities or currents, which can be interpreted as a kind of "generalized wave packets". As a natural basis for unperturbed states of the zero approximation in the interaction of an electron with an electromagnetic field and with other particles in the density matrix method for the open system under consideration, we take the complete system of one-electron stationary wave functions. These wave functions below the vacuum level describe electronic states limited by the volume inside the photocathode, and above the vacuum level they are solutions to the problem of electron scattering by the volume and surface potential and describe delocalized states propagating inside and outside the photocathode. In the presence of a thin quantum size heterostructure, the last wave functions contain preexponential coefficients proportional to the scattering amplitudes, which can have pole singularities, providing a resonant oscillatory contributions of quasi-stationary states to the probability and current densities, both directly and through the spectrum of the density matrix elements. 

The second purpose of this article is numerical-analytical study of a one-dimensional model of the mechanism of wave-like modulation of a photoelectron pulse during its passage through a double-well quantum-well heterostructure deposited on a volumetric planar photocathode. The wavelike spatiotemporal modulation of the pulse of the charge density and the current density of photoelectrons arises as a result of excitation by this pulse and the subsequent slow oscillatory decay of the doublet of quasi-stationary states of the heterostructure. We will show that it is possible to provide prolongation of generation and even amplification of waves of charge density and current density of photoelectrons when the photocathode is exposed to a periodic sequence of light pulses, such that their durations and intervals between them are multiples of the difference period of the doublet.

\section{THE EMISSION CHARGE AND CURRENT DENSITIES. STATEMENT OF THE PROBLEM AND THE CHOICE OF MODEL}

The charge $n(\bm r,t)$  and current  $\bm j(\bm r,t)$ densities at a point $\bm r$  at a time $t$  are given \cite{Abri1975} by universal expressions
\begin{equation*}
n(\bm r,t) =  - ie G\left({t,\bm r_0 ;t + 0, \bm r} \right)_{\bm r_0  = \bm r} 
\end{equation*}
\begin{equation*}
{\bm{j}}(\bm r,t)\! =\! \frac{{e\hbar }}
{m}\!\left( {\nabla _{\bm r} G\!\left( {t,\bm r_0 ;t + 0,\bm r} \right)\! -\! \nabla _{\bm r} G\!\left( {t,\bm r;t + 0,\bm r_0 } \right)} \right)_{\bm r_0  = \bm r} 
\end{equation*}
where $G\left( {t_1 ,\bm r_1; t_2 ,\bm r_2 } \right) = i\left\langle {\hat \Psi ^ +  (t_2, r_2 )\hat \Psi (t_1 ,r_1 )} \right\rangle$ is two-time causal Green$'$s function, $\hat \Psi (r,t) = \sum\nolimits_p {\hat a_p (t)\psi _p (r)} $, 
$\hat a_p (t)$  is the Heisenberg field operator, $\psi_p (\bm r)$
  is two-time causal Green's function,  
Schrodinger wave function of an electron in a stationary state  $p$,  $\left\langle  \ldots  \right\rangle$ is the statistical average over the equilibrium state of an unperturbed system. 
Let us denote by  $
\hat \rho _{p_1 ,p_2 ,} (t) = \hat a_{p_1 }^ +  (t){\kern 1pt} {\kern 1pt} \hat a_{p_2 }^{} (t + 0)$
 the operator of the two-time density matrix at coinciding times $t$ , and by $\rho _{p_1 ,p_2 } (t) \equiv \left\langle {\hat \rho _{p_1 ,p_2 ,} (t)} \right\rangle $
  the matrix elements of the density matrix. We introduce the time-independent "matrix elements" $n_{p,p'} (\bm r)$
  and  ${\mathbf{j}}_{p,p'} (\bm r)$ 
 \cite{Land1977} of the charge $\hat n(\bm r)$  and current ${\mathbf{\hat j}}(\bm r)$
 densities at the point  $\bm r$
\begin{equation}\label{eq:math:1}
n_{p,p'} (r) = e\psi _{p'}^* (r)\psi _p (r)
\end{equation}
\begin{equation}\label{eq:math:2}
{\mathbf{j}}_{p,p'} (r) = i\frac{{e\hbar }}
{{2m}}\left[ {\left( {\nabla \psi _{p'}^* (r)} \right)\psi _p (r) - \psi _{p'}^* (r)\left( {\nabla \psi _p (r)} \right)} \right].
\end{equation}

\begin{figure}[h]
\includegraphics[width=7.5 cm]{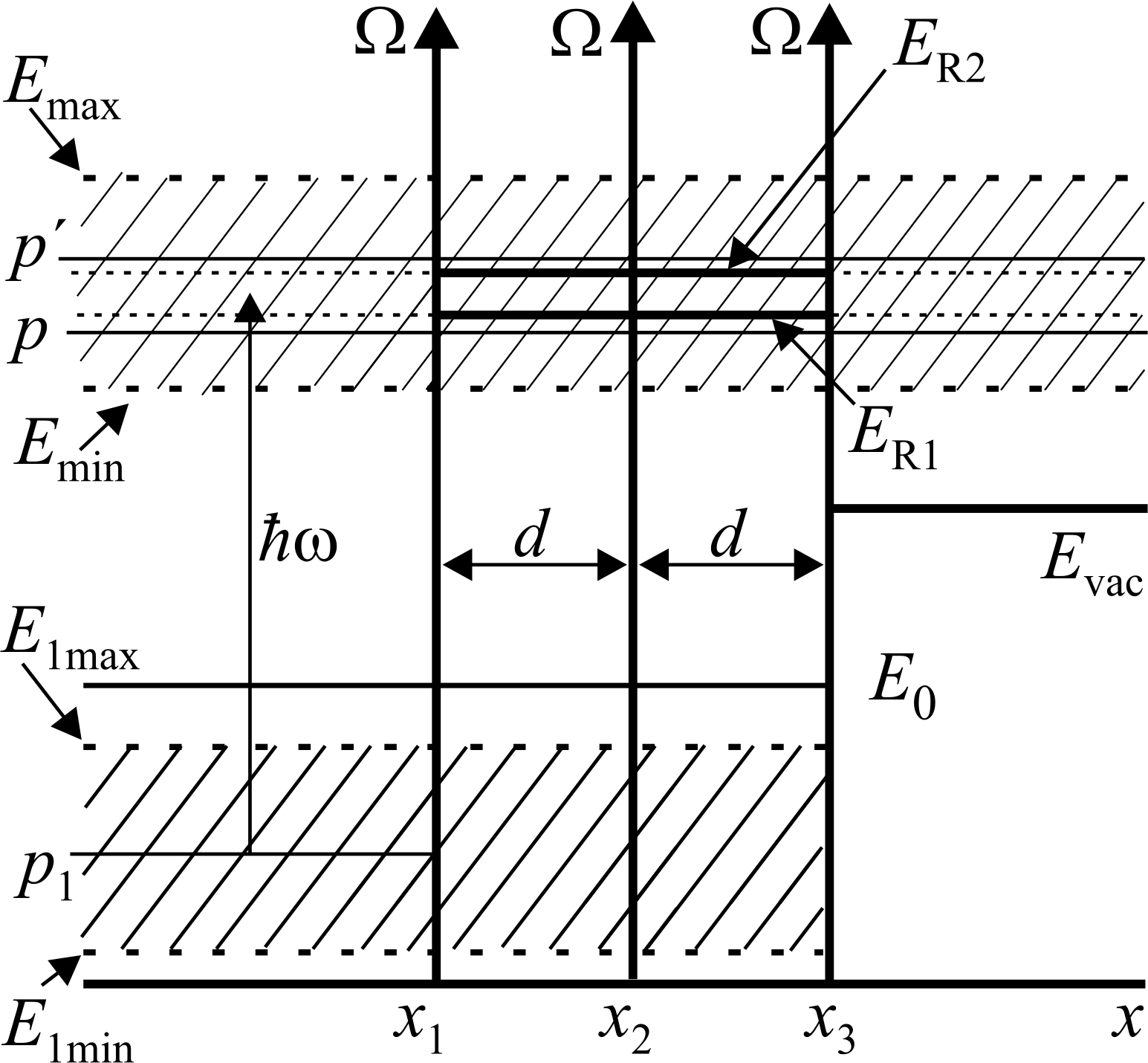}\\
  \caption{ Coordinate-energy diagram of a photoemitter with a surface heterostructure.}\label{FIG:Fig1}
\end{figure}

These quantities are not statistical, but microscopic. For fast processes that occur in a time much shorter than the time required to establish thermodynamic equilibrium, they can be removed from the sign of statistical averaging, and the charge and current densities can be represented as
\begin{equation}\label{eq:math:3}
n(\bm r,t) = 2\;Sp\left( {\hat \rho (t)\hat n(\bm r)} \right) = 2\sum\limits_{p,p'} {\rho _{p',p} (t)n_{p,p'} (\bm r)} 
\end{equation}
\begin{equation}\label{eq:math:4}
{\mathbf{j}}(\bm r,t) = 2\;Sp\left( {\hat \rho (t){\mathbf{\hat j}}(\bm r)} \right) = 2\sum\limits_{p,p'} {\rho _{p',p} (t){\mathbf{j}}_{p,p'} (\bm r)} 
\end{equation}
The latter expressions obviously generalize to an open system of rigorous expressions for the charge $n_c(\bm r,t)$  and current   $\mathbf{j}_c(\bm r,t)$ densities in the "pure" quantum-mechanical state of the wave packet type
\begin{equation}\label{eq:math:5}
\Psi_c\equiv\Psi_c(\bm r,t)=\sum_pc_pe^{-iEt}\psi_p(\bm r)
\end{equation}
where  $E=E(p)$   energy of an electron in a stationary state  $p$,   $c_p$  the spectral function:
\begin{equation*}
n_c (\bm r,t) \equiv |\Psi _c |^2  = \sum\limits_{p,p'} {\rho _{p',p}^c n_{p,p'} (\bm r)} 
\end{equation*}
\begin{equation}\label{eq:math:6}
\begin{gathered}
\mathbf{j}_c (\bm r,t) \equiv i\frac{{e\hbar }}
{{2m}}\left[ {\left( {\nabla \Psi _c^* } \right)\Psi _c  - \Psi _c^* \left( {\nabla \Psi _c } \right)} \right] =  \hfill \\
  \quad \quad \quad \quad \quad \quad \quad \quad \quad \quad  = 2\sum\limits_{p,p'} {\rho _{p',p}^c } (t)\mathbf{j}_{p,p'} (\bm r) \hfill \\ 
\end{gathered} 
\end{equation}
\begin{equation*}
\rho _{p',p}^c (t) = \rho _{p',p}^{c*} (t) = c_p c_{p'}^* e^{ - i(E - E')t}. 
\end{equation*} 
With the main goal of extracting the discussed resonance contributions to the photocurrent normal to the surface, in this article we will consider the quasi-one-dimensional model depicted in the coordinate-energy diagram (Fig.\ref{FIG:Fig1}), when a heterostructure in the form of a double quantum well formed by three identical tunnel-transparent potential barriers  $\Omega$ at a distance  $d$ from each other is deposited on the flat surface of a bulk photocathode, the heterostructure plays the role of an energy filter for photoelectrons.

When photoexcited by light of frequency  $\omega$, electrons undergo transitions between states $p_1$  below the vacuum level, localized inside the photocathode and delocalized above vacuum states  $p$ and $p'$   (Fig.\ref{FIG:Fig1}). The detector of the photocurrent normal to the surface must be located outside the system and must be configured to register the discussed alternating current of photoemission of electrons with energies $\varepsilon _p $   and  $\varepsilon _p' $
 from a narrow band  $E_{\min }  \leqslant \varepsilon _p ,\varepsilon _{p'}  \leqslant E_{\max } $
 of states $p$  and  $p'$, covering one doublet of resonant quasi-stationary states with energies $E_{R1}$  and $E_{R2}$  above the vacuum level of the photocathode. Because of the law of conservation of energy, by light of a given frequency  $\omega$ electrons will effectively be excited into the states of such a band  $E_{\min }  \leqslant \varepsilon _p ,\varepsilon _{p'}  \leqslant E_{\max } $, the initial states   of which belong to some also narrow energy band $E_{1\min }  \approx E_{\min }  - \hbar \omega  \leqslant \varepsilon _{p_1 }  \leqslant E_{\max }  - \hbar \omega  \approx E_{1\max }   $
  below the vacuum level $E_{vac}$  and the boundary level  $E_0$ (for a metal photocathode, this is the Fermi energy of a partly filled conduction band, and for a semiconductor photocathode this is the energy of the valence band ceiling). 

Here we are interested in the pulses of the charge and current densities of photoexcited electrons, which are equal to the sums \eqref{eq:math:3} and \eqref{eq:math:4} over the states $p$  and $p'$  in the continuous spectrum of the scattering problem, belonging to a narrow registered energy band of photoelectrons. The wave functions of electrons $\psi_p(\bm r)$  are the basic solutions of the stationary Schrodinger equation, which takes into account the spatial profile of the potential energy of the electron. The elements of the density matrix  $\rho_{p',p}(t)$ obey the kinetic equation and carry information about the photoexcitation of electrons from deep-lying stationary states, as well as about inelastic scattering processes. Comparison of expressions 
\eqref{eq:math:1}
 - \eqref{eq:math:4} and \eqref{eq:math:6} shows that the nonstationary pulses of the charge and current densities arising as a result of photoexcitation and scattering of electrons, retain and generalize the most important properties of the wave packet \eqref{eq:math:5} formed by the superposition of the wave functions of excited stationary states of electrons. In the previous article \cite{Peis2021}, we studied in detail the case when relations \eqref{eq:math:5} - \eqref{eq:math:6} describe the scattering of a Gaussian wave packet by a double-well heterostructure of the Fig.\ref{FIG:Fig1} type and showed that outside the heterostructure $n_c(\bm r,t)$  and  $\mathbf{j}(\bm r,t)$ undergo wave-like space-time modulation. When a short photoemission pulse is scattered by such a heterostructure, similar effects should also appear.

\section{SOLUTION OF THE KINETIC EQUATION FOR THE DENSITY MATRIX}
The density matrix operator obeys the equation of motion \cite{Ross2002}, \cite{Kope1970}-\cite{Ilyi1989}
\begin{equation}\label{eq:math:7}
i\hbar \frac{\partial }
{{\partial t}}\hat \rho _{p_1 ,p_2 }  = [\hat H,\hat \rho _{p_1 ,p_2 } ]
\end{equation}
the Hamiltonian of the system has the form \cite{Nabu1976}
\begin{equation}\label{eq:math:8}
\hat H = \sum\limits_p {\xi _p \hat a_p^ +  \hat a_p }  - \sum\limits_{p_1 ,p_2 } {Ed_{p_1 ,p_2 } \hat a_{p_1 }^ +  \hat a_{p_2 }^{} }  + \hat H_1,
\end{equation}
 where $\xi _p  = \varepsilon _p  - \mu $
is the energy of the electron in the state  $p$,  $\mu$ is the chemical potential, $\hat H_1 $  is the part of the Hamiltonian describing the electron-electron and electron-phonon interaction, it leads to a renormalization of energy levels, that is, to their shift $\Delta E_p$  and smearing  $\gamma_p$. If the system is acted upon by a pulse of light pumping of a characteristic duration  $t_0$, then the electric field strength of an electromagnetic wave can be represented as a Fourier expansion $E(t) = \sum\nolimits_\omega  {E_\omega  (t)e^{i\omega t} } $, where  $\omega  =  \pm |\omega |$
 are high light frequencies  $(|\omega | \gg t_0^{ - 1} )$, and the envelopes $E_\omega  (t)$  are functions of time with a scale of change  $t_0$, for simplicity we will simulate them with rectangular pulses of duration $t_0$  (Fig. 2) ,
\begin{equation}\label{eq:math:9A}
\bm d_{p_1 ,p_2 }  = \int {\psi _{p_1 }^ *  ({\mathbf{r}}){\kern 1pt} {\kern 1pt} e{\mathbf{r}}\psi _{p_2 } ({\mathbf{r}})} d^3 {\mathbf{r}}
\end{equation}
are matrix elements of the electron electric dipole moment. Opening the commutator and averaging in the mass operator   
$M_p  = \Delta \varepsilon _p  + i\gamma _p $
 (or relaxation time $\hbar \gamma _p^{ - 1} $) approximation, we obtain a system of kinetic equations for the elements of the density matrix
\begin{equation}\label{eq:math:9}
\hbar \frac{\partial }
{{\partial t}}\rho _{p',p}  - i\xi _{p,p'} \rho _{p',p}  = \hbar F_E \left\{ \rho  \right\} - \gamma _{p,p'} \left( {\rho _{p',p}  - \bar \rho _{p',p} } \right)
\end{equation}
where 
\[
F_E \left\{ \rho  \right\} = \frac{i}
{\hbar }\sum\limits_{p_1 } {\left\{ {\left( {Ed_{p_1 ,p} } \right)\rho _{p',p_1 }  - \left( {Ed_{p',p_1 } } \right)\rho _{p_1 ,p} } \right\}} 
\]
and  $\xi _{p,p'}  = \xi _p  - \xi _{p'} $ -   the difference between the renormalized energies, $\gamma _{p,p'}  = \gamma _{p'}  + \gamma _p  > 0$ -  the total width of the combined levels.  The Hermiticity of the density matrix $
\rho _{p',p}  = \rho _{p,p'}^ * $  is ensured by the fact that  $\xi _{p,p'}  =  - \xi _{p',p} $
 and  $\gamma _{p,p'}  = \gamma _{p',p} $. In equilibrium, only the diagonal elements of the density matrix on the initially occupied states with Fermi average occupation numbers $n_p$  are not equal to zero \cite{Ilyi1989}
\begin{equation}\label{eq:math:10}
\rho _{p',p}^{(0)}  = \bar \rho _{p',p}  = n_p \delta _{p',p}  = \left\{ {\begin{array}{*{20}c}
   {n_p ,} & {p' = p}  \\
   {0\quad ,} & {p' \ne p}  \\
 \end{array} } \right.
\end{equation}
Further developing the theory of perturbation in the electric field $\bm E=\bm E(t)$ 
\[
\rho  = \rho ^{(0)}  + \rho ^{(1)}  + \rho ^{(2)}  +  \ldots,\;\rho ^{(n)}\!  \sim \!E^n,\; \Omega _R\! =\! Ed/\hbar  \ll | {\omega _{p'p} } |,
\]
we have a system of recurrent differential equations
\begin{equation}\label{eq:math:11}
\frac{\partial }
{{\partial t}}\rho _{p',p}^{(n)}  - i\omega _{p',p} \rho _{p',p}^{(n)}  = F_{p',p}^{(n)} (t),\quad \quad n = 0,1,2,...
\end{equation}
where
\[
F_{p',p}^{(0)} (t) = \frac{1}
{\hbar }\gamma _{p,p} n_p \delta _{p',p} ,\quad 
\]
and for $n \geq 1$
\[
F_{p',p}^{(n)} (t){\kern 1pt} {\kern 1pt}  = {\kern 1pt} {\kern 1pt} \frac{i}
{\hbar }\sum\limits_{p_1 } {\left\{ {\left( {Ed_{p_1 ,p} } \right)\rho _{p',p_1 }^{(n - 1)}  - \left( {Ed_{p',p_1 } } \right)\rho _{p_1 ,p}^{(n - 1)} } \right\}},
\]
and $\omega _{p',p}  = \left( {\xi _{p,p'}  + i\gamma _{p',p} } \right)/\hbar $, moreover $\omega _{p,p'}  =  - \omega _{p',p}^ *$. 
The general solution to each of equations 
\eqref{eq:math:11} 
has the form (for $n \geq 1$)
\begin{equation}\label{eq:math:12}
\rho _{p',p}^{(n)} (t) = e^{i\omega _{p',p} t} \left\{ {\int\limits_0^t {F_{p',p}^{(n)} (\tau )e^{ - i\omega _{p',p} \tau } d\tau }  + \rho _{p',p}^{(n)} (0)} \right\}.
\end{equation}

We are interested in transient processes for times on the order of the relaxation times of the electronic subsystem when the light pulse is suddenly switched on and off. At each recurrent step, integration over time gives
\begin{equation}\label{eq:math:13}
\int\limits_0^t {E(\tau )e^{ - i\omega _{p_1 ,p_2 } \tau } d\tau  \approx i\sum\limits_\omega  {\frac{{E_\omega  \left( t \right)}}
{{\omega _{p_1 ,p_2 }  - \omega }}} } \left( {e^{i(\omega  - \omega _{p_1 ,p_2 } )t}  - 1} \right),
\end{equation}
where in brackets the exponent oscillating with time is the contribution of the upper limit of integration, and the subtracted unit is the contribution of the lower limit, that is, the moment of switching on. When describing stationary photoemission, the light is usually considered to be strictly monochromatic from a certain moment in time, but its switching-on adiabatically moves away from $t=0$  to  $t=-\infty$  (for example, by introducing an infinitely slowly increasing time exponent until the moment of stabilization), in this case, in integrals like 
\eqref{eq:math:12}, the contribution of the lower limit becomes zero, that is, formally in brackets \eqref{eq:math:13}, one should be replaced by zero. In the case of short light pulses under consideration, the contribution of the moment when the light is switched on is significant, and the contribution of the upper limit of integration is given by terms proportional $\exp (i\omega t)$.
\begin{figure}[h]
\includegraphics[width=7.5 cm]{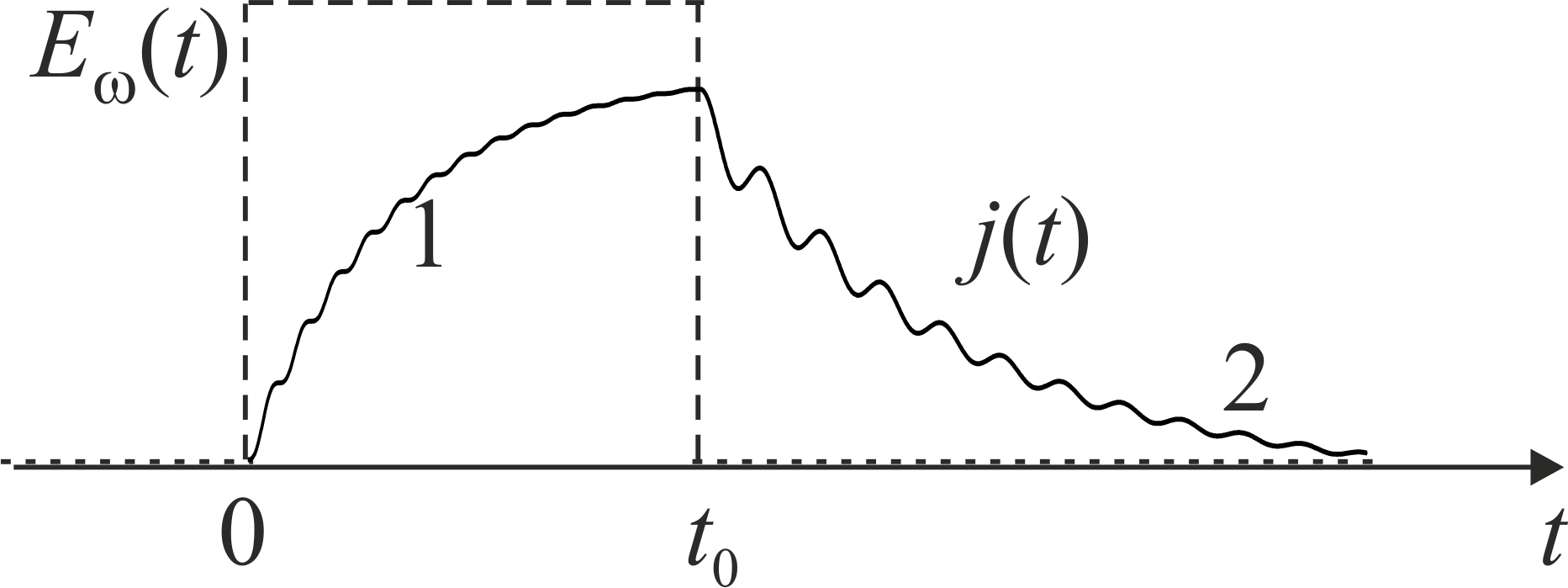}\\
  \caption{The response of the photocurrent (solid line) as a function of time against the background of a rectangular pump pulse (dashed line) for a system like Fig.\ref{FIG:Fig1}.}\label{FIG:Fig2}
\end{figure} 
 
If at the initial moment $t=0$  the off-diagonal elements of the density matrix are equal to zero $\rho _{p',p}^{(n)} (0) = 0$, then at $t>0$  the first order solution in the electric field has the form
\begin{equation}\label{eq:math:14}
\rho _{p',p}^{(1)} (t) = \frac{1}
{\hbar }\left( {n_{p'}  - n_p } \right)\sum\limits_\omega  {\left( {d_{p',p} E_\omega  } \right)\frac{{e^{i\omega t}  - e^{i\omega _{p',p} t} }}
{{\omega  - \omega _{p',p} }}}, 
\end{equation}
and the second order solution is given by the expression
\begin{equation}\label{eq:math:15}
\begin{gathered}
  \rho _{p',p}^{(2)} (t) =  \hfill \\
   = \frac{1}
{{\hbar ^2 }}\sum\limits_{p_1 ,\omega ,\omega _1 } {\left\{ {\frac{{\left( {n_{p'}  - n_{p_1 } } \right)\left( {d_{p',p_1 } E_\omega  } \right)\left( {d_{p_1 ,p} E_{\omega _1 } } \right)}}
{{\omega  - \omega _{p',p_1 } }}} \right.}  \times  \hfill \\
   \times \left. {\left[ {\frac{{e^{i\left( {\omega  + \omega _1 } \right)t}  - e^{i\omega _{p',p} t} }}
{{ {\omega  + \omega _1  - \omega _{p',p} } }} + } \right.\frac{{e^{i\left( {\omega _{p',p_1 }  + \omega _1 } \right)t}  - e^{i\omega _{p',p} t} }}
{{ {\omega _{p',p}  - \omega _{p',p_1 }  - \omega _1 } }}} \right] +  \hfill \\
   + \frac{{\left( {n_p  - n_{p_1 } } \right)\left( {d_{p',p_1 } E_{\omega _1 } } \right)\left( {d_{p_1 ,p} E_\omega  } \right)}}
{{\omega  - \omega _{p_1 ,p} }} \times  \hfill \\
  \left. { \times \left[ {\frac{{e^{i\left( {\omega  + \omega _1 } \right)t}  - e^{i\omega _{p',p} t} }}
{{ {\omega  + \omega _1  - \omega _{p',p} } }} + \frac{{e^{i\left( {\omega _{p_1 ,p}  + \omega _1 } \right)t}  - e^{i\omega _{p',p} t} }}
{{ {\omega _{p',p}  - \omega _{p_1 ,p}  - \omega _1 } }}} \right]} \right\}, \hfill \\ 
\end{gathered} 
\end{equation}
and here the formal frequency parameters $\omega$  and  $\omega_1$ take positive and negative values.

It is essential in the same way that the photocurrent does not stop instantly after a sharp switch-off of the light pulse. If at some moment the  $t=t_0$ off-diagonal elements of the density matrix $\rho _{p',p}^{(2)} (t)$
  reach values $\rho _{p',p}^{(2)} (t_0)$  and at this moment the exciting light pulse is abruptly switched off  $\bm E(t)=0$, then, in accordance with \eqref{eq:math:12}, for  $t>t_0$, the solution of Eq. \eqref{eq:math:11} has the form
\begin{equation}\label{eq:math:16}
\rho _{p',p}^{(2)} (t) = e^{i\omega _{p',p} (t - t_0 )} \rho _{p',p}^{(2)} (t_0 )
\end{equation}
where $\rho _{p',p}^{(2)} (t_0)$  is calculated by formula \eqref{eq:math:15}, i.e.  $\rho _{p',p}^{(2)} (t)$ oscillates and decays exponentially over time.

\section{ANALUSIS OF THE MAIN CONTRIBUTIONS}
When calculating the photoelectron charge and current densities, substituting \eqref{eq:math:15} or \eqref{eq:math:16}  into \eqref{eq:math:3} and \eqref{eq:math:4}, one should neglect small terms. The states $p$  and $p'$  that determine the current are initially not occupied, but excited by light, and have equilibrium values of the Fermi occupation numbers, which are practically zero $n_{p'}  \approx n_p  \approx 0$, while the unexcited initially occupied states $p_1$  have occupation numbers almost equal to unity $n_{p_1 }  \approx 1$. Under excitation by almost monochromatic light with a frequency  $|\omega|$, the denominators in \eqref{eq:math:15} have such a structure that the terms with $\omega _1  =  - \omega $
  are large compared to other terms that can be neglected, moreover, for the large terms in the first square bracket  $\omega  < 0,\;\omega _1  > 0$, and in the second square bracket  $\omega  > 0,\;\omega _1  < 0$, therefore we leave only them and further, denoting  $\omega  = |\omega | = |\omega _1 | > 0$, we have
\begin{equation}\label{eq:math:17}
\begin{gathered}
  \rho _{p',p}^{(2)} (t) = \sum\limits_{p_1 } {D_{p_1 } }  \times  \hfill \\
  \left[ {\frac{{e^{i\omega _{p',p} t}  - 1}}
{{\omega _{p',p} \left( {\omega  + \omega _{p',p_1 } } \right)}} + \frac{{e^{i\omega _{p',p} t}  - e^{i\left( {\omega _{p',p_1 }  + \omega } \right)t} }}
{{\left( {\omega  + \omega _{p',p_1 } } \right)\left( {\omega  + \omega _{p',p_1 }  - \omega _{p',p} } \right)}}} \right. \hfill \\
  \left. { + \frac{{1 - e^{i\omega _{p',p} t} }}
{{\omega _{p',p} \left( {\omega  - \omega _{p_1 ,p} } \right)}} + \frac{{e^{i\omega _{p',p} t}  - e^{i\left( {\omega _{p_1 ,p}  - \omega } \right)t} }}
{{\left( {\omega  - \omega _{p_1 ,p} } \right)\left( {\omega  - \omega _{p_1 ,p}  + \omega _{p',p} } \right)}}} \right], \hfill \\ 
\end{gathered} 
\end{equation}
here and below, $D_{p_1}$  denote the coefficients
\begin{equation}\label{eq:math:18}
D_{p_1 }  = \frac{{n_{p_1 } }}
{{\hbar ^2 }}\left( {d_{p',p_1 } E_{ - \omega } } \right)\left( {d_{p_1 ,p} E_\omega  } \right),
\end{equation}
they are proportional to the product of the moduli of the matrix elements of the electron dipole moments $| {\bm d_{p',p_1 } } || {\bm d_{p_1 ,p} }|$  and the light intensity  $| {E_\omega  } |^2 $; therefore, they essentially determine the magnitude of the photoelectron charge and current densities.

Let us express the difference frequencies in terms of the energies and damping decrements of the stationary states combined by them $
\hbar \omega _{p_1 ,p_2 }  = \xi _{p_2 ,p_1 }  + i\gamma _{p_1 ,p_2 } = \;\xi _{p_2 }  - \xi _{p_1 }  + i\gamma _{p_1 ,p_2 } $. We also take into account that the smearing of high-energy excited states usually exceeds the smearing of unexcited states $\gamma _p  \sim \gamma _{p'}  \gg \gamma _{p_1 }$, then \eqref{eq:math:17} takes the form
\begin{equation}\label{eq:math:19}
\begin{gathered}
  \rho _{p',p}^{(2)} (t) = \frac{{\hbar ^2 }}
{{(\xi _p  - \xi _{p'} ) + i\gamma _{p'p} }}{\kern 1pt} \left\{ {\left[ {1 + e^{i(\xi _p  - \xi _{p'} ){t \mathord{\left/
 {\vphantom {t \hbar }} \right.
 \kern-\nulldelimiterspace} \hbar } - \gamma _{p'p} {t \mathord{\left/
 {\vphantom {t \hbar }} \right.
 \kern-\nulldelimiterspace} \hbar }} } \right]} \right.\!\!\! \times  \hfill \\
  \sum\limits_{p_1 } {D_{p_1 } } f(\omega ,p,p',p_1 ) - {\kern 1pt} \sum\limits_{p_1 } {D_{p_1 } } f(\omega ,p,p',p_1 ) \times  \hfill \\
  \left. {\left[ {e^{i\left( {\hbar \omega  - (\xi _{p'}  - \xi _{p_1 } )} \right){t \mathord{\left/
 {\vphantom {t \hbar }} \right.
 \kern-\nulldelimiterspace} \hbar } - \gamma _{p'} {t \mathord{\left/
 {\vphantom {t \hbar }} \right.
 \kern-\nulldelimiterspace} \hbar }}\!\!  +\!\! e^{ - i\left( {\hbar \omega  - (\xi _p  - \xi _{p_1 } )} \right){t \mathord{\left/
 {\vphantom {t \hbar }} \right.
 \kern-\nulldelimiterspace} \hbar } - \gamma _p {t \mathord{\left/
 {\vphantom {t {\hbar }}} \right.
 \kern-\nulldelimiterspace} {\hbar }}} } \right]} \right\}, \hfill \\ 
\end{gathered} 
\end{equation}
where 
\begin{equation}\label{eq:math:20}
\begin{gathered}
  f(\omega ,p,p',p_1 ) =  \hfill \\
   = \frac{1}
{{\hbar \omega  - (\xi _p  - \xi _{p_1 } ) - i\gamma _p }} - \frac{1}
{{\hbar \omega  - (\xi _{p'}  - \xi _{p_1 } ) + i\gamma _{p'} }}. \hfill \\ 
\end{gathered} 
\end{equation}

All time-dependent terms are related to the contributions of the lower limits of integration over time; they reflect the influence of the moment of switching on the light pulse and oscillate with frequencies close to  $(\xi _p  - \xi _{p'} )/\hbar$, decaying exponentially in times of the order of relaxation times $\gamma _{p',p}^{ - 1}  \sim \;\gamma _{p'}^{ - 1}  \sim \gamma _p^{ - 1} $. At $t\to \infty$  each of the quantities $\rho _{p',p}^{(2)} (t)$  tends to a constant value determined by the contributions of the upper limits of integration over time (the unit inside the square bracket of the first term):
\begin{equation}\label{eq:math:21}
\rho _{p',p(0)}^{(2)}  = {\kern 1pt} {\kern 1pt} \frac{{\hbar ^2 }}
{{(\xi _p  - {\kern 1pt} \xi _{p'} ) + {\kern 1pt} i\gamma _{p'p} }}\sum\limits_{p_1 } {D_{p_1 } } f(\omega ,p,p',p_1 ).
\end{equation}	 
Substitution $\rho _{p',p(0)}^{(2)}$  instead of  $\rho _{p',p}^{(2)} (t)$  in \eqref{eq:math:3} and \eqref{eq:math:4} gives expressions for the charge and current density of stationary photoemission in the energy band recorded by the detector.

	        The general expressions of Sections 3 and 4 derived in the relaxation time approximation are applicable to a photocathode in which the processes of inelastic scattering of electrons are weak, that is, the thickness of the region of photoexcitation of electrons is less than the mean free path of high-energy electrons emitted into vacuum immediately after photoexcitation in a pulsed fast one-step quantum coherent process. This can be a bulk photocathode located at  $x<x_0$, within which unexcited states $p_1$  are localized, and the heterostructure is absent or located on the surface (Fig.\ref{FIG:Fig1}). Even better, our general formulas are applicable to the description of photoemission from a separate double quantum well, which is a thin-film photocathode whose thickness is less than the mean free path of electrons. 
	        If the photocurrent is formed by pulsed photoexcitation of electrons directly in thin conducting layers from the inside of a quantum double-well heterostructure, then the pole features of the scattering amplitudes of excited states of electrons $\psi_p$  and  $\psi_{p'}$ should manifest themselves not only explicitly through expressions \eqref{eq:math:1} and \eqref{eq:math:2}, but also through the matrix elements of the dipole moments \eqref{eq:math:9A}. Our preliminary calculations show that in this case the effect of quasi-wave beats and modulation of the charge and current densities going in both directions from such a photocathode can be stronger. This issue requires additional research.
	        
	          Formula \eqref{eq:math:21} does not contradict the formulas of the one-step model of photoemission \cite{Caro1973}, which express the almost coherent quasi-elastic part of the photocurrent, which is proportional to the sum of the products of three dressed Keldysh$'$s Green$'$s functions; in our case, they correspond to the factors
$G^R  \sim (\xi _p  - \xi _{p'}  + i\gamma _{p'p})^{-1}$, 
$G^A  \sim (\hbar \omega  -\xi _{p'}  + \xi _{p_1 }  \mp i\gamma _p )^{-1}$, 
$G^ +   \sim n_{p_1 }  $.

General microscopic \cite{Caro1973} and phenomenological three-step \cite{Spic1958,Spic1993} theories of photoemission indicate that in bulk photocathodes, the thickness of which is much greater than the mean free path of excited electrons, the quasi-elastic approximation is insufficient, and the magnitude of the photocurrent is strongly influenced by the processes of multiple inelastic scattering of electrons mainly by phonons. This slow stage of the process, which contributes to the accumulation of excited electrons at the bottom of the conduction band before they escape into vacuum (which provides a large photocurrent), is often described by the equations of diffusion theory \cite{Hart1999,Aule2002} (in thick photocathodes, especially in photocathodes with negative electron affinity). Our consideration is not applicable to such cases.  
  
	          Our formulas are convenient in that they contain easily interpretable characteristics of the energy spectrum of electrons and light; they are valid for the fast stage of the process, as long as the light field is not too strong and the scattering by phonons and electrons is rather weak.
	          
	        It can be seen from them that at $\gamma\to 0$  and  $p'\approx p$, the main contribution to stationary photoemission is associated with the products of two blurry  $\delta$-functions describing the approximate conservation of energy upon photoexcitation of an electron: the optical (interband) Joint density of states per unit phase volume
\[
\delta (\xi _{p_1 }  + \hbar \omega {\kern 1pt} {\kern 1pt}  - \xi _{p'} ) \approx \frac{1}
{\pi }\frac{{\gamma _{p'p_1 } }}
{{(\xi _{p_1 }  + \hbar \omega {\kern 1pt} {\kern 1pt}  - \xi _{p'} )^2  + \gamma _{p'p_1 }^2 }},
\]
and the intraband density of excited states per unit phase volume
\[
\delta (\xi _p  - \xi _{p'} ) \approx \frac{{\gamma _{p'p} }}
{{{\kern 1pt} {\kern 1pt} {\kern 1pt} (\xi _p  - \xi _{p'} ){\kern 1pt} ^2 {\kern 1pt}  + {\kern 1pt} \gamma _{p'p}^2 }}.
\]

In this approximation, the expression for the steady-state current corresponds to a phenomenological three-step model of photoemission with allowance for weak blur of states \cite{Nabu1976}. The distribution of photoelectrons over states with energies $\xi _p $
   is given by the sub-sum \eqref{eq:math:4} over states  $p'$. Sometimes, for an estimate, it is assumed that the main contribution to the total photocurrent comes from terms \eqref{eq:math:4}) with  $p'=p$, however, the terms of sum \eqref{eq:math:4} with off-diagonal terms $p'\neq p$  and real parts of the Lorentzian fractions from \eqref{eq:math:19} - \eqref{eq:math:21} can also make a noticeable contribution to the photocurrent, especially in its variable part.

\section{SIMPLIFIED CALCULATION FORMULAS}

 When calculating the resonant photocurrent of interest to us through the double quantum well (or from the well) using formulas \eqref{eq:math:3}, \eqref{eq:math:4} we must sum over the excited states  $p$ and  $p'$, which belong to a narrow band $E_{\min }  \leqslant \varepsilon _p ,\varepsilon _{p'}  \leqslant E_{\max } $  of energies recorded by the detector and which covers above-vacuum doublet of mutually close quasi-stationary levels with energies $E_{R1}$  and  $E_{R2}$, the distance between these levels is small in comparison with the width of the allowed energy bands of the photocathode participating in the optical transition, and both  $E_{\min}$ and   $E_{\max}$ are also far from other quasi-stationary levels.
 
	            In this case, the calculation $\rho _{p',p}^{(2)}(t)$  by formulas \eqref{eq:math:19} and \eqref{eq:math:21} requires summation over the initial unexcited states $p_1$  that belong to a certain energy strip $E_{1\min }   \leqslant \varepsilon _{p_1 }  \leqslant    E_{1\max }$ (where $E_{1\min }<E_{\min }-\hbar\omega$, $E_{1\max }>E_{\max }-\hbar\omega$) in the partially filled conduction band or in the valence band of the photocathode, for which the resonance denominators in expression \eqref{eq:math:20} are sufficiently small.
            
            The absolute values of the photocurrent and the quantum yield of photoemission can vary over a very wide range depending on the fundamental frequency and intensity $\left| {E_\omega  } \right|^2 $  of light, as well as on the physico-chemical nature of the photocathode material and the structure of potential barriers. In the theory of photoemission from bulk photocathodes (as in the theory of  light absorption and reflection spectra), the intensity of the light electric field  $E_\omega  $, as well as the matrix elements of the dipole moment  $\bm d_{p',p_1 } , \bm d_{p_1 ,p} $ (i.e., parameters  $D_{p_1 }  \approx D$), can be considered almost constant factors in the corresponding ranges.
            
           It is convenient to replace the summation over the states in formulas \eqref{eq:math:3}, \eqref{eq:math:4} and \eqref{eq:math:19}, \eqref{eq:math:21} by numerical integration over energies, introducing factors equal to the energy densities of states $g_p  = dN/d\varepsilon _p $
  in a certain volume. Stationary wave functions should also be normalized in the same volume. If $L$  is the normalization length along the axis  $x$, then $\psi (E,x) \sim 1/\sqrt{ L }$,  $g_p\sim L$, and in \eqref{eq:math:3} and \eqref{eq:math:4} the dependence on $L$  is canceled.
           
	          In this paper, we are not interested in the threshold and saddle singularities of the densities of states; and for points of general position in narrow bands of width $\Delta E$  within the allowed energy bands of the photocathode, the quantities  $g_p, \;\; g_{p'}$ and  $g_{p_1}$ can be considered as constants of the order  $g_p\sim\Delta N/\Delta E$, where  $\Delta N$ is the number of electronic states in the band  $\Delta E$; for the same reason, we can neglect the energy dependence of the damping parameters  $\gamma_p$. In any case, we perform photoemission calculations of the space-time dependences of the charge and current densities up to an unknown constant factor associated with normalization, light intensity, and values of the matrix elements of the optical transition. For simplicity, you can take  $E_\omega  ,g_p ,g_{p'} ,g_{p_1 } $,  and $D_{p_1 }  \approx D$
  equal to units (if necessary, these factors can be estimated numerically). Specifically, we calculated the dimensionless ratios of the photoemission charge $n(x,t)$  and current $j(x,t)$  densities to their maximum values in the absence of a heterostructure for such a narrow energy band that these almost constant factors were reduced.

            Obviously, the spectral width, duration, and length of the photocurrent pulse increase with increasing width of the summation energy interval  $[ {E_{\min }, E_{\max } } ]$. We performed calculations using formulas \eqref{eq:math:3}, \eqref{eq:math:4}, substituting in them the results of summation over $p_1$  in expressions \eqref{eq:math:19} and \eqref{eq:math:21}. Such calculations show that if the widths of the energy bands  $[ {E_{\min } , E_{\max } } ]$, $[ {E_{1\min } , E_{1\max } } ]$ are large enough compared to the distance between the resonance levels   and   of the doublet of quasi-stationary states, then the difference spatiotemporal component of the modulated photoemission pulse of interest to us is qualitatively and quantitatively not very sensitive to the choice of boundaries $[ {E_{\min } ,E_{\max } } ]$  and $[ {E_{1\min } , E_{1\max } } ]$  within wide limits. Therefore, under these conditions, it is possible with acceptable accuracy to calculate the main resonance contribution to the integrals, which express the sums over $p_1$  in expressions \eqref{eq:math:19} and \eqref{eq:math:21} as shown in Appendix.
            
	         This makes it possible to write down rather simple expressions for the elements of the density matrix instead of \eqref{eq:math:19} for the pumping process:
\begin{equation}\label{eq:math:23}
\rho _{p',p}^{(2)} (t) = \frac{{2\pi i\hbar ^2 D}}
{{(\xi _p  - \xi _{p'} ) + i\gamma _{p'p} }}{\kern 1pt} \left[ {1 - e^{i(\xi _p  - \xi _{p'} ){t \mathord{\left/
 {\vphantom {t \hbar }} \right.
 \kern-\nulldelimiterspace} \hbar } - \gamma _{p'p} {t \mathord{\left/
 {\vphantom {t \hbar }} \right.
 \kern-\nulldelimiterspace} \hbar }} } \right]
\end{equation}
instead of \eqref{eq:math:21} after entering the stationary saturation mode:
\begin{equation}\label{eq:math:24}
\rho _{p',p(0)}^{(2)}  = \frac{{2\pi i\hbar ^2 D}}
{{(\xi _p  - \xi _{p'} ) + i\gamma _{p'p} }}
\end{equation}
and instead of  \eqref{eq:math:16} at $t>t_0$ after switching off the pumping:
\begin{equation}\label{eq:math:25}
\rho _{p',p}^{(2)} (t) = \rho _{p',p}^{(2)} (t_0 )e^{i(\xi _p  - \xi _{p'} ){{(t - t_0 )} \mathord{\left/
 {\vphantom {{(t - t_0 )} \hbar }} \right.
 \kern-\nulldelimiterspace} \hbar } - \gamma _{p'p} {{(t - t_0 )} \mathord{\left/
 {\vphantom {{(t - t_0 )} \hbar }} \right.
 \kern-\nulldelimiterspace} \hbar }}, 
\end{equation}
where $\rho _{p',p}^{(2)} (t_0 )$  is the initial value arbitrarily set at the moment  $t_0$, which can be estimated by expressions \eqref{eq:math:23} or \eqref{eq:math:24}. Note that these expressions did not include the frequency of light  $\omega$ due to the rapid convergence of integrals \eqref{eq:math:A2}, which approximate the sums \eqref{eq:math:19} and \eqref{eq:math:21}.

           Substitution of \eqref{eq:math:23} - \eqref{eq:math:25} into \eqref{eq:math:3} and \eqref{eq:math:4} gives practically the same oscillation-relaxation dependence of the photoelectron charge density and current density on time (of the type (Fig.\ref{FIG:Fig2})) and coordinates as substitution of \eqref{eq:math:19} - \eqref{eq:math:21}.

\section{NUMERICAL SIMULATION OF PULSE PHOTOCURRENT}

We wish to demonstrate the manifestation of the resonance contributions of the poles of the scattering amplitudes of photoelectrons by a double quantum well to the sums describing the nonstationary photoemission current normal to the surface of a planar photocathode. For this, when constructing the wave functions of emitted electrons, we restrict ourselves to the simplest quasi-one-dimensional model (Fig.\ref{FIG:Fig1}) of the Sommerfeld model type, replacing the lattice potential acting on these electrons with the potential of a rectangular barrier with a height  $E_{vac}$  at  $x=x_3$  (the axis $x$  is directed across the surface of the photocathode and heterostructure). The bottom of such a potential is determined by the electron affinity $\chi$  in the photocathode crystal; for simplicity, in the calculations, we will assume it to be the same in the conducting layers of the heterostructure, the potential barriers of which will be modeled by three delta functions $U(x) = \left( {\hbar ^2 /2m} \right)\sum\nolimits_{n = 1}^3 {\Omega \delta \left( {x - x_n } \right)} $  of the same power   $\Omega$ at a distance $d$  from each other at $x_1=0,\;x_2=d\;x_3=2d$. Delta barriers can be used to model real, fairly narrow and high potential barriers, in this case, the estimate $\Omega  \approx 2mU_b d_b /\hbar ^2 $ is valid, where  $U_b$ is the height of the barrier,  $d_b$ is its width. The energy of electrons will be measured from the vacuum level.

\begin{figure}[h]
\includegraphics[width=8.5 cm]{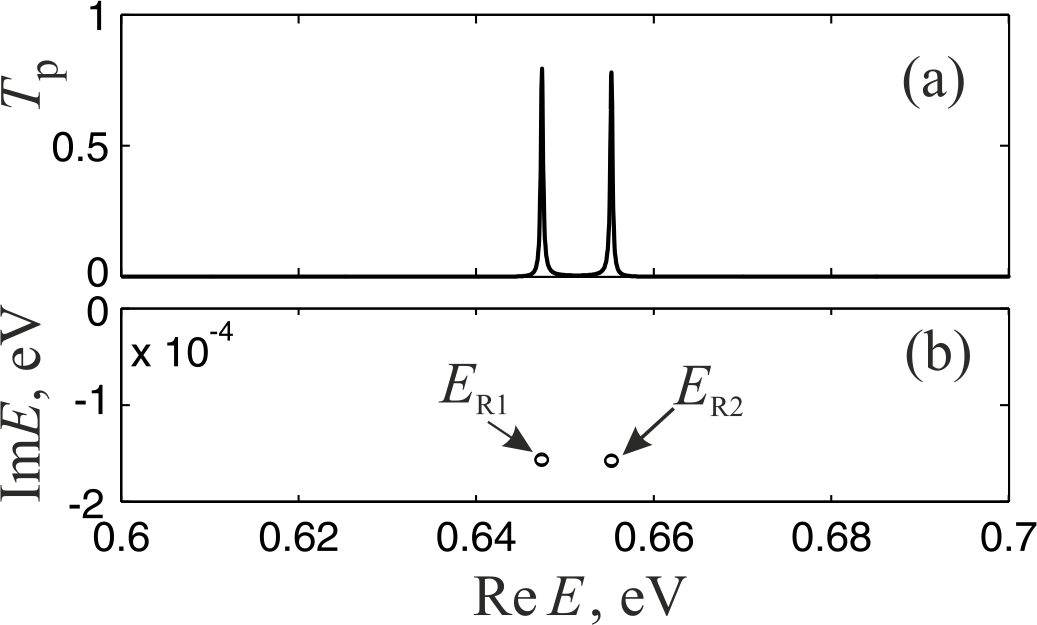}\\
  \caption{  The studied doublet of a) the transmission coefficient $T_p$  and b) the poles of the transmission amplitude $t_p$  through the heterostructure.}\label{FIG:Fig3}
\end{figure}

Thus, we assume that the required wave functions of excited electrons to the left and right of the heterostructure have approximately the form
\begin{equation}\label{eq:math:26} 
\psi _p (r) = \frac{1}
{{\sqrt L }}\left\{ {\begin{array}{*{20}c}
   {e^{ik_{p_0 } x}  + r_p e^{ - ik_{p_0 } x} ,} & {x \leqslant x_1 }  \\
   {t_p e^{ik_{p_3 } \left( {x - x_3 } \right)} ,} & {x \geqslant x_3 }  
 \end{array} } \right.
\end{equation}
where $k_{p_0 }  = \hbar ^{ - 1} \sqrt {2m(E + \chi )} $
  is the quasi-wave number transverse to the boundary in the volume of the photocathode, $k_{p_3 }  = \hbar ^{ - 1} \sqrt {2mE} $ is the wave number in vacuum to the right of the system, $r_p$  and $t_p$  are the amplitudes of reflection and transmission of the surface barrier with a heterostructure, $L$   is the normalization length. Substituting the second row of \eqref{eq:math:26} into \eqref{eq:math:1} and \eqref{eq:math:2}, we have expressions for the matrix elements of the charge and current densities at points  $x  \geqslant  x_3 $ outside the heterostructure
\begin{equation}\label{eq:math:27} 
\begin{gathered}
  n_{p'p} \left( {x} \right) = \frac{e}
{L}t_{p'} t_p^* e^{i(k_{p'_3 }  - k_{p_3 } )(x - x_3 )}  \hfill \\
  j_{p'p} \left( {x} \right) = \frac{{e\hbar }}
{{2mL}}t_{p'} t_p^* \left( {k_{p'_3 }  + k_{p_3 } } \right)e^{i(k_{p'_3 }  - k_{p_3 } )(x - x_3 )}  \hfill \\ 
\end{gathered} 
\end{equation}
where  $k_{p'_3 }  = \hbar ^{ - 1} \sqrt {2mE'}$. The transmission amplitude $t_p$  can be found analytically or numerically using its expression through the elements of the effective transfer matrix $M_{ef}$ \cite{Peis2008A, Peis2008B, Peis2021} by the formulas
\begin{equation}\label{eq:math:28} 
\begin{gathered}
  t_p  = \frac{{\det M_{ef} }}
{{\left( {M_{ef} } \right)_{22} }},\quad \quad M_{ef}  = L_3^{ - 1} M_\Omega  MM_\Omega  MM_\Omega  L_0 , \hfill \\
  M = \left( {\begin{array}{*{20}c}
   {\cos kd} & {\frac{{\sin {\kern 1pt} \,kd}}
{k}}  \\
   { - k\sin kd} & {\cos \,kd}  \\

 \end{array} } \right),\; \hfill \\
  M_\Omega   = \left( {\begin{array}{*{20}c}
   1 & 0  \\
   \Omega  & 1  \\

 \end{array} } \right),\;\quad L_j  = \left( {\begin{array}{*{20}c}
   1 & 1  \\
   {ik_{p_j } } & { - ik_{p_j } }  \\

 \end{array} } \right), \hfill \\ 
\end{gathered} 
\end{equation}
where  $k = k_{p_0 }  = \hbar ^{ - 1} \sqrt {2m(E + \chi )} $
 is the quasi-wave number in the heterostructure, $j=0$  or $j=3$. The quantities $t_p$, $n_{p'p}$, and $j_{p'p}$  have pole singularities in the lower half-plane of the complex energy of the electron at the values of the complex energy $E_R  =  \operatorname{Re} E_R  + i\operatorname{Im} E_R $
  determined by the equality to zero of the matrix element 
  $(M_{ef})_{22}$, these poles are associated with the position of the narrow peaks of the transmission coefficient through the heterostructure $T_p=(k_{p_3}/k_{p_0})|t_p|^2$  (Fig.\ref{FIG:Fig3}). The quantities $\operatorname{Re} E_R $
  give the energies of quasi-stationary states in the heterostructure, which are approximately equal to the energies of the peaks $T_p$, and the quantities $ - \operatorname{Im} E_R  \equiv \Gamma _R $
  give the widths of the peaks  $T_p$, as well as the energy widths of the quasi-stationary states and their lifetimes  $\tau _R  = \hbar /\Gamma _R $ \cite{Peis2008A,Peis2008B,Peis2021}.
  
		Below we present the results of numerical simulation for a photocathode with a surface heterostructure, a simplified energy diagram of which is shown in Fig.\ref{FIG:Fig1}, for the following specified parameters: $d=125$ \AA,  $\Omega=10$ a.u.$=18.9$ \AA$^{-1}$, $\chi=4$  eV.  By solving numerically the equation   $(M_{ef})_{22}=0$, we establish that the doublets lower above the vacuum level are located near energies (0.035 eV, 0.042 eV), (0.234 eV, 0.242 eV), (0.439 eV, 0.446 eV), (0.647 eV-0.655 eV), (0.861 eV , 0.869 eV), ... . Difference oscillations of the densities of the photoemission charge and current can be manifested by a "wave packet" formed by a superposition of photoelectrons with energies from a certain band $E_{\min }  \leqslant E \leqslant E_{\max } $, which is wide enough to cover one doublet of resonant quasi-stationary states, but narrow compared to the distances to neighboring doublets. Such a pulse can be created by separating photoelectrons with energies $E_{\min }  \leqslant E \leqslant E_{\max } $
  through the use of magnetic and electric fields of the appropriate configuration.
  
		We have calculated the densities of the photoemission charge and current generated by excited electrons, the energies of which belong to the band enclosing the fourth supra-vacuum doublet, which corresponds to two mutually close poles of the transmission amplitude $t_p$  (i.e., the roots of the equation $(M_{ef})_{22}=0$: $E_{R1}  = (0.647 - i1.567 \cdot 10^{ - 4} )$  eV and $E_{R2}  = (0.655 - i1.576 \cdot 10^{ - 4} )$
  eV.  Fig.\ref{FIG:Fig3} b) shows the position of this doublet on the complex energy plane, and Fig.\ref{FIG:Fig3}, a) shows the spectrum of the transmission coefficient through the heterostructure.

It is seen that the heterostructure is practically impenetrable outside resonances, and the narrow resonance peaks of the transparency coefficient $T_p$  have a width of the order of the imaginary part of the poles. 	For the lifetimes of quasi-stationary states associated with this doublet, we have values  
$\tau _{R1}  \approx \hbar /|\operatorname{Im} E_{R1} | = 4.18 \cdot 10^{ - 12} $ s $ = 1.73 \cdot 10^5 $  a.u., $\tau _{R2}  \approx \hbar /|\operatorname{Im} E_{R2} |=4.16 \cdot 10^{ - 12} $
 s $ = 1.72 \cdot 10^5 $ a.u., that is $\tau _{R1}  \approx \tau _{R2} $. The difference between the energies of these states  $\Delta E_{R12}  = \operatorname{Re} E_{R2}  - \operatorname{Re} E_{R1}  = 0.0078$ eV determines the frequency $\nu _{12}  = \Delta E_{R12} /2\pi \hbar  = 1.89 \cdot 10^{13} $ Hz and the period $T_{12}  = 1/\nu _{12}  = 5.29 \cdot 10^{ - 13} $ s  $ = 2.2 \cdot 10^4 $ a.u. oscillations of the photocurrent.
 
Oscillations of the current will be effectively observed when the inequalities  $\tau _{R1} ,\tau _{R2} ,\tau _p  \gg T_{12} $
 are satisfied, where $\tau _p  = \hbar /\gamma _p $
  is the electron relaxation time determined by inelastic scattering. In numerical calculations, we used the value $\gamma _p  = 2.72 \cdot 10^{ - 5} $
 eV, i.e. $\tau _p  = 2.4 \cdot 10{}^{ - 11}$ s $ = 1 \cdot 10^6 $ a.e., that is typical for bulk semiconductors.
 
	In expressions \eqref{eq:math:3} and \eqref{eq:math:4}, we pass from the summation over the numbers of states   and   to the integration over the energies of these states   and  :
\begin{equation}\label{eq:math:29} 
n(x,t) = 2\iint\limits_S {\rho _{p',p} (t)n_{p,p'} (x)}g_p g_{p'} dEdE',
\end{equation}
\begin{equation}\label{eq:math:30} 
j(x,t) = 2\iint\limits_S {\rho _{p',p} (t)j_{p'p} (x)}g_p g_{p'} dEdE',
\end{equation}
here $\rho _{p',p} (t)$  is given \eqref{eq:math:23} or \eqref{eq:math:25} with $\xi _p  - \xi _{p'}  = E - E'$
  and $\gamma _{p'p}  = 2\gamma _p  = \rm{const}$; $n_{p,p'} (x)$
  and $j_{p,p'} (x)$  are given \eqref{eq:math:27} with $k_{p_3 }  = \hbar ^{ - 1} \sqrt {2mE}$,   and $k_{p'_3 }  = \hbar ^{ - 1} \sqrt {2mE'}$  from \eqref{eq:math:28}; integration is performed over the square $S$  in which $E_{\min }  \leqslant E,E' \leqslant E_{\max } $.

In the calculations, we took the boundaries of the detected energy band to be equal to $E_{\min }  = 0.63$
 eV and $E_{\max }  = 0.67$ eV. Due to the rapid convergence of integrals \eqref{eq:math:29} and \eqref{eq:math:30}, the oscillatory contribution of the poles to the calculated charge  $n(x,t)$ and current $j(x,t)$  densities is almost independent of the choice of these boundaries in a wide enough range between neighboring doublets, although the absolute values of  $n(x,t)$ and  $j(x,t)$ increase with increasing of integration bandwidth. As mentioned above, we calculated the dimensionless ratios of the photoemission densities of charge  $n(x,t)$ and current $j(x,t)$  to their maximum values for the same photocathode without a heterostructure  $n_0$ and $j_0$. This reduces the dependences on specific values $I_{\omega}, g_p ,g_{p'} ,g_{p_1 } $,  and $D_{p_1 }  \approx D$  because they are almost constant values in narrow bands of integration  $E_{\min }  \leqslant E,E' \leqslant E_{\max } $.	In contrast to analogous integrals corresponding to sums \eqref{eq:math:6} for charge $n_c(x,t)$  and current $j_c(x,t)$  densities in a "pure" quantum mechanical state of the wave packet type \eqref{eq:math:5}, the double integrals \eqref{eq:math:29} and \eqref{eq:math:30} cannot be expressed in terms of the product of two single integrals of the type of the integral corresponding to the sum \eqref{eq:math:5} due to the energy denominator  $(\xi _p  - \xi _{p'} ) + i\gamma _{p'p}  = (E - E') + i2\gamma _p $. Therefore, asymptotic estimates of these integrals by the fastest descent method \cite{Peis2021} are difficult.

\begin{figure}[h]
\includegraphics[width=6.5 cm]{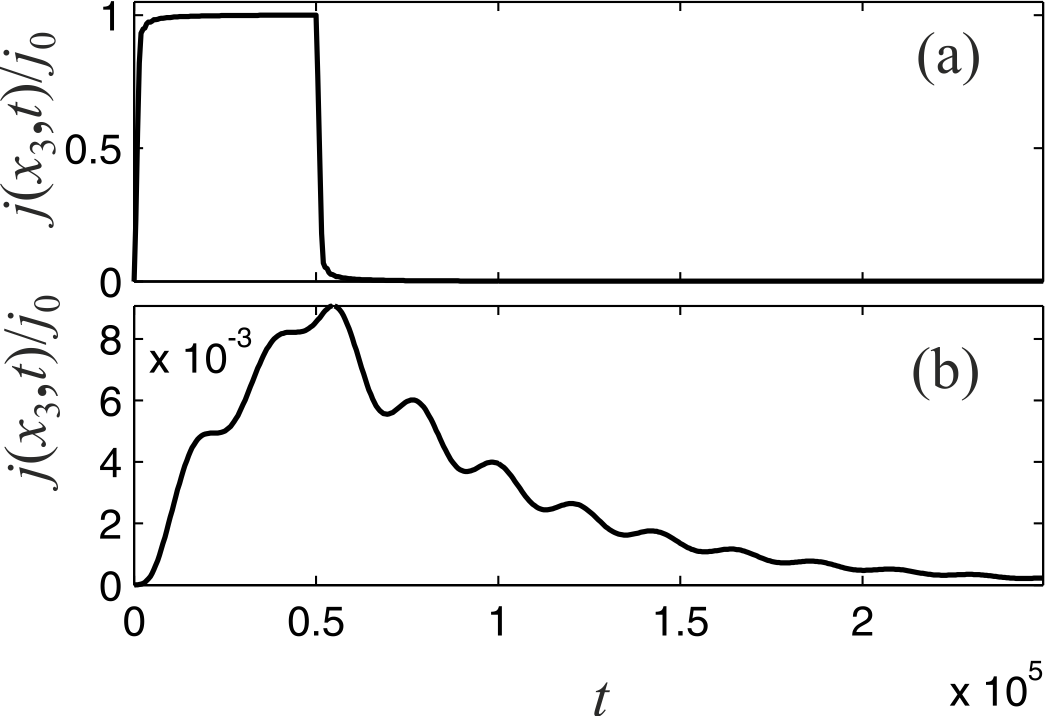}\\
  \caption{Time dependence of the current density pulse at the exit point $x_3=2d$  for cases (a) the absence of a heterostructure, (b) the presence of a heterostructure. Time in atomic units  1 a.u. (t) $ = 2.419 \times 10^{ - 17} $ s.}\label{FIG:Fig4}
\end{figure}

\begin{figure}[h]
\includegraphics[width=6.5 cm]{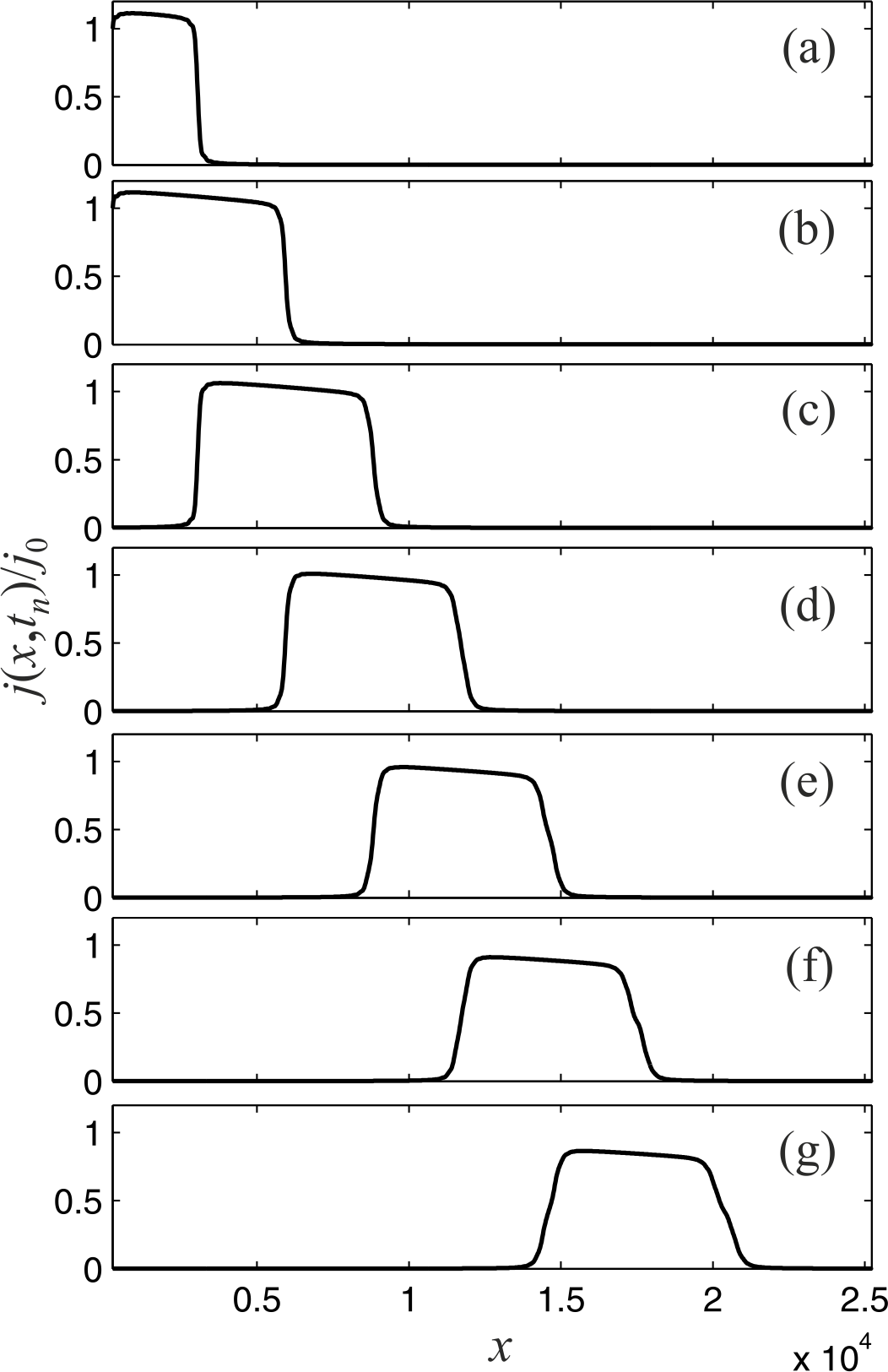}\\
  \caption{Coordinate dependence of the current density pulse from a photocathode without a surface heterostructure for different instants of time  $t_n$: (a) $t_1  = 2.5 \cdot 10^4 $
  a.u., (b)  $t_2  = 5 \cdot 10^4 $
 a.u., (c) $t_3  = 7.5 \cdot 10^4 $
  a.u., (d) $t_4  = 10 \cdot 10^4 $
  a.u., (e)  $t_5  = 12.5 \cdot 10^4 $
 a.u., (f) $t_6  = 15.0 \cdot 10^4$ a.u., (g) $t_7  = 17.5 \cdot 10^4 $
  a.u. Coordinate $x$  in angstroms \AA.}\label{FIG:Fig5}
\end{figure}

\begin{figure}[h]
\includegraphics[width=6.5 cm]{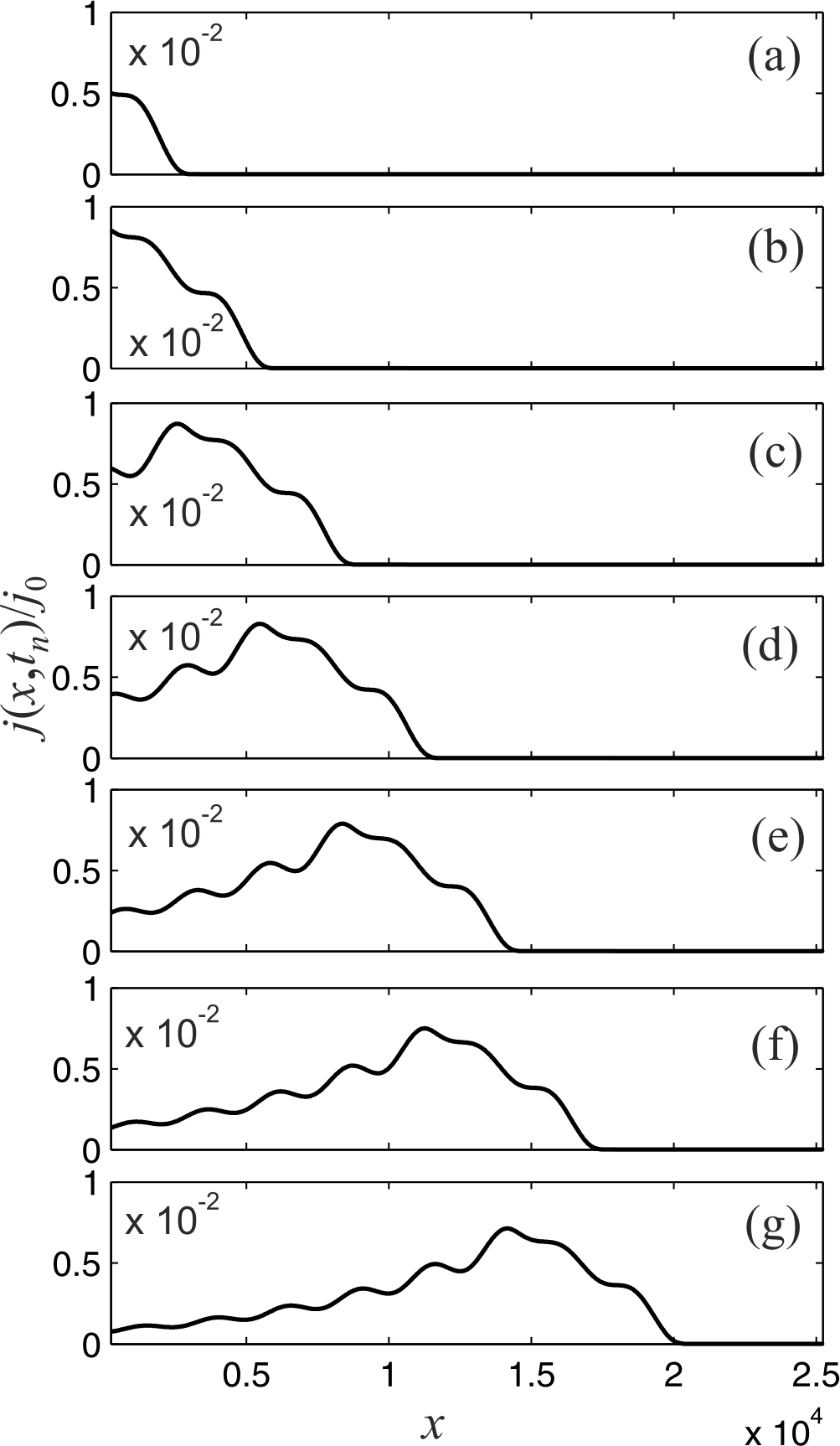}\\
  \caption{  Coordinate dependence of the current density pulse from a photocathode with a surface heterostructure for different instants of time  $t_n$: (a) $t_1  = 2.5 \cdot 10^4 $
  a.u., (b)  $t_2  = 5 \cdot 10^4 $
 a.u., (c) $t_3  = 7.5 \cdot 10^4 $
  a.u., (d) $t_4  = 10 \cdot 10^4 $
  a.u., (e)  $t_5  = 12.5 \cdot 10^4 $
 a.u., (f) $t_6  = 15.0 \cdot 10^4$ a.u., (g) $t_7  = 17.5 \cdot 10^4 $
  a.u. Coordinate $x$  in angstroms \AA.}\label{FIG:Fig6}
\end{figure}

We have obtained the sought space-time dependences of the photoelectron charge densities  $n(x,t)$ and  current densities   $j(x,t)$ by direct numerical integration of expressions \eqref{eq:math:29} and \eqref{eq:math:30}. Subsequent figures Fig.\ref{FIG:Fig4} - Fig.\ref{FIG:Fig6}. demonstrate these dependences for a pulse of the photocurrent density   at the duration of a rectangular pumping light pulse $t_0  = 1.21 \cdot 10^{ - 12} $ s $ = 5 \cdot 10^4 $ a.u. 
Similar figures for these dependences of the pulse of the charge density $n(x,t)$  look qualitatively almost the same, this is obvious from a comparison of two expressions \eqref{eq:math:27}: in contrast to $n_{p'p}(x)$, the quantity  $j_{p'p}(x)$ contains a factor $
{{\hbar (k_{p'_3 }  + k_{p_3 } )} \mathord{\left/
 {\vphantom {{\hbar (k_{p'_3 }  + k_{p_3 } )} {2m}}} \right.
 \kern-\nulldelimiterspace} {2m}}$
  that hardly changes within a narrow integration band $E_{\min }  \leqslant E \leqslant E_{\max } $.
  
In the absence of a heterostructure, i.e. at  $\Omega=0$, $d=0$ the amplitude of the transmission of a rectangular step is
\[t_p  = \frac{{2k_{p_0 } }}
{{k_{p_0 }  + k_{p_3 } }},\]
in this case, the time dependence of the current density pulse at the point of exit from the heterostructure  $x_3=2d$ has the form (Fig.\ref{FIG:Fig4}, a), as $j_0$  we took the maximum value of the current density at this point.

	In the presence of a heterostructure in the form of a double quantum well on the photocathode surface, the time dependence of the photocurrent density pulse at the exit point from the heterostructure $x_3=2d$  varies greatly and has the form (Fig.\ref{FIG:Fig4} b). After switching off the light pulse, it is strongly extended in time, demonstrating a slow exponential decay over a time interval $ \sim \tau _{R1}  \approx \tau _{R2}  =  \cdot 10^{ - 11 \pm 1} $ s $ = 10^{6 \pm 1} $ a.u. and temporal oscillations with a period $ \approx T = 5.3 \cdot 10^{ - 13} $ s $ = 2.2 \cdot 10^4 $ a.u. close to the period  $T_{12}$ of the difference frequency of the selected doublet. Oscillations of this kind occur both during light pumping up to the instant $t_0$  and after the instant $t_0$  of switching off the light pulse in the process of slow relaxation decay of quasi-stationary states.

    A rough estimate of the points of stationarity of the phases of the integrands \eqref{eq:math:29} and \eqref{eq:math:30} in two variables $E$  and $E'$  indicates that the pulses $n(x,t)$  and $j(x,t)$  should move along  $x$ with a velocity close to the group velocity of the wave packet  ${\text{v}}_g  = \hbar ^{ - 1} {{\partial E} \mathord{\left/
 {\vphantom {{\partial E} {\partial k = }}} \right.
 \kern-\nulldelimiterspace} {\partial k = }}{{\hbar k} \mathord{\left/
 {\vphantom {{\hbar k} m}} \right.
 \kern-\nulldelimiterspace} m}$, where $\hbar k = \sqrt {2mE_c } $  approximately corresponds to the spectral center  $E_c$ of the packet, which gives ${\text{v}}_g  \approx 4.8 \cdot 10^5 $ m / s for $E_c  = 0.65$  eV.
 
The coordinate dependence of the current density pulse from a photocathode without a surface heterostructure for different instants of time is shown in (Fig.\ref{FIG:Fig5}). After formation, over a period of time  $t_0$, a pulse with a length of about $\Delta x \approx {\text{v}}_g t_0  \approx 0.58 \cdot 10^4 $ \AA ~moves with a speed of about  $v_g$, experiencing weak damping and smearing.
	
If there is a heterostructure in the form of a double quantum well on the photocathode surface, the coordinate dependence of the photocurrent density pulse for different instants of time is shown in (Fig.\ref{FIG:Fig6}). One can see spatial oscillations with a period length $\lambda  = {{2\pi } \mathord{\left/
 {\vphantom {{2\pi } {\left| {k_{R2}  - k_{R1} } \right|}}} \right.
 \kern-\nulldelimiterspace} {\left| {k_{R2}  - k_{R1} } \right|}} \approx 2544$ \AA  ~corresponding to the difference in wave numbers $
k_{R2}  = \operatorname{Re} \left( {\hbar ^{ - 1} \sqrt {2mE_{R2} } } \right)
$
  and $
k_{R1}  = \operatorname{Re} \left( {\hbar ^{ - 1} \sqrt {2mE_{R1} } } \right)
$, determined by resonant quasi-stationary levels $E_{R2}$  and  $E_{R1}$. Oscillations are present both on the leading edge formed during pumping and on the long tail formed during the slow decay of quasi-stationary states in the quantum well, which decays exponentially over a length of  $
\Delta x \sim  - \hbar \left( {\operatorname{Im} \sqrt {2mE_{R1} } } \right)^{ - 1}  \approx  - \hbar \left( {\operatorname{Im} \sqrt {2mE_{R2} } } \right)^{ - 1}  \approx 10^5 
$ \AA.

Comparison of figures Fig. \ref{FIG:Fig3}(a) and Fig. \ref{FIG:Fig3}(b) (as well as Figures \ref{FIG:Fig5} and \ref{FIG:Fig6}) shows that in the presence of a surface heterostructure with the selected parameters $\Omega=10$ a.u. $=18.8$ \AA$^{-1}$ and $d=125$ \AA  ~the maximum value of the photocurrent pulse is approximately two orders of magnitude lower than in the absence of the heterostructure, due to the low transparency of the potential barriers of the heterostructure. The wavelike space-time oscillations of the photocurrent with a difference frequency  $\nu _{12}  = (\operatorname{Re} E_{R2}  - \operatorname{Re} E_{R1} )/2\pi \hbar $, period  $T_{12}  = 1/\nu _{12} $, and wavelength   $\lambda _{12}  = 2\pi /|k_{R2}  - k_{R1} |$ 
 are obviously associated with the manifestation in integrals \eqref{eq:math:29} and \eqref{eq:math:30} of two pairs of narrow stripes, on which the energies are close to the values    $E = E_{R1} ,\,{\kern 1pt} E = E_{R2}$ and $
E' = E_{R1} ,\,{\kern 1pt} E' = E_{R2} $
 of pole features of the amplitude of transmission through the surface double well  $t_p$. At the same time, a narrow stripe in which the energies are close to the values satisfying $
(\xi _p  - \xi _{p'} ) + i\gamma _{p'p}  = (E - E') + i2\gamma _p  = 0$
 (for which the energy denominator $\rho _{p',p} (t)$  is singularly small) together with the full width of the integration region, determine the magnitude of the charge density and current density pulses components, which are smooth in coordinate and time. In the absence of a double quantum well on the photocathode surface, this smooth component completely describes the photocurrent. In the presence of a surface double well, it is also not small, but the oscillatory integral contributions of the poles  $t_p$ may well appear on its background and compete with it.

\section{PROLONGATION AND AMPLIFICATION OF WAVE GENERATION}

The process of generating the quasi-wave component of the photoelectronic charge and current densities with the difference frequency $\nu_{12}$  and wavelength $\lambda_{12}$  of the doublet of quasi-stationary states of a double quantum well located on the surface of the photocathode can be continued and even amplified, if the photocathode is illuminated with a sequence of identical quasi-rectangular pulses, the duration of which $t_0=nT_{12}$  and the interval between which $\delta t=sT_{12}$  are multiples (i.e. $n$  and $s$   are natural numbers) of the difference period $T_{12}$  of the doublet. This corresponds to the second method considered in \cite{Peis2021} for creating a sequence of almost identical pulse wave packets prepared in one place, here in the region of the surface heterostructure sequentially in time with a time period  $\delta t$, as a result of this, coherent wave impulses of $n(x,t)$  and $j(x,t)$  of the form Fig.\ref{FIG:Fig4} (b) and Fig.\ref{FIG:Fig6}. will move to the right one after the other with overlapping. The sequence of $N$  such pulses can provide prolongation or even amplification (up to  $N$-fold at  $s=0$, and to a lesser extent at  $s=1,2,3,...$) oscillating pulses. 

\begin{figure}[h]
\includegraphics[width=6.5 cm]{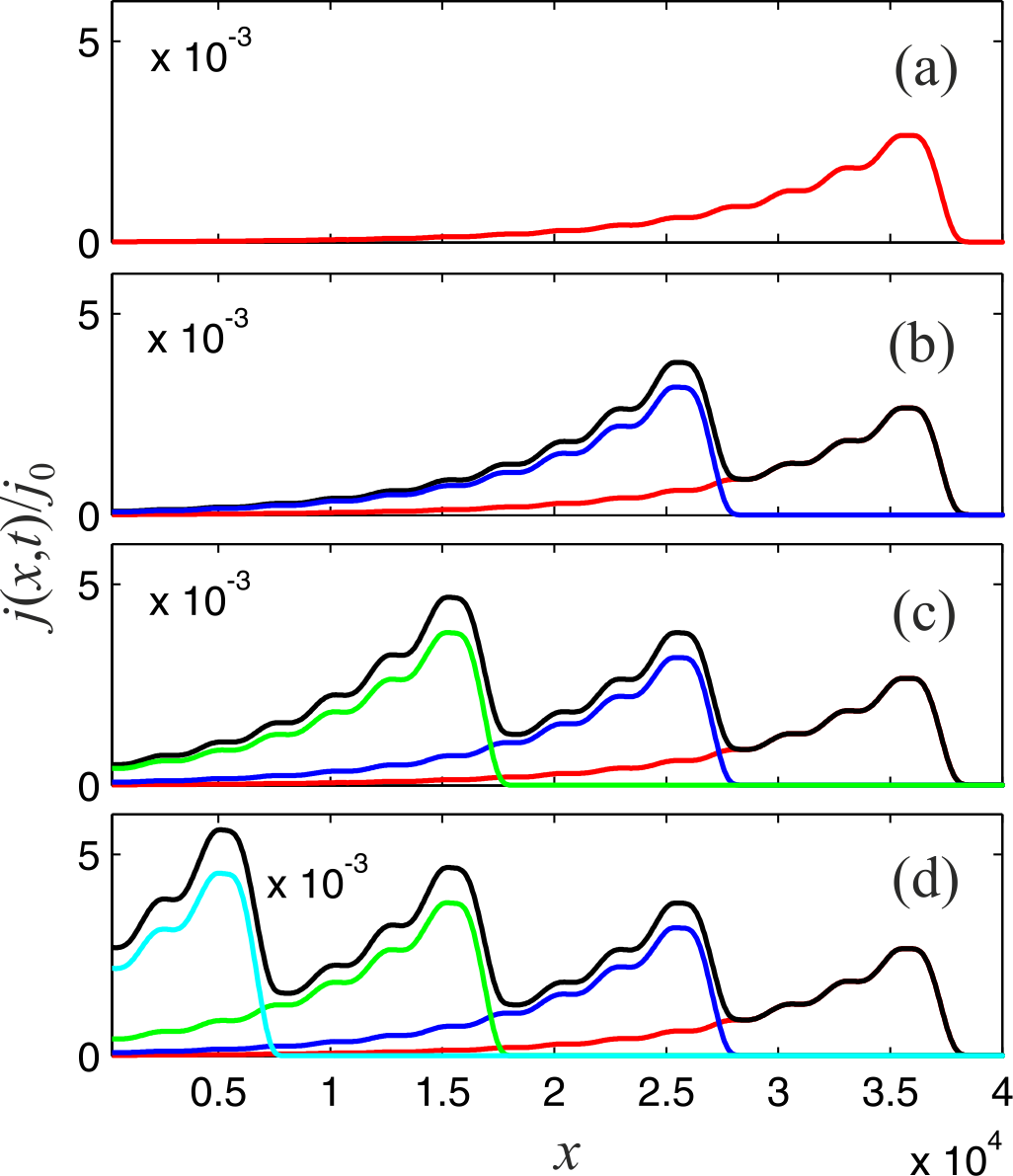}\\
  \caption{(Color online) Time profile of the photocurrent density at the point   $x_3=2d$ of exit from the heterostructure as a result of the action of (a) one and sequences (b) two, (c) three, (d) four identical pump light pulses with a duration $t_0=T_{12}$  shifted in time by  $\delta t=4T_{12}$. Time in atomic units  ($t$) a.u. $ = 2.419 \times 10^{ - 17} $ s.}\label{FIG:Fig7}
\end{figure}

\begin{figure}[h]
\includegraphics[width=6.5 cm]{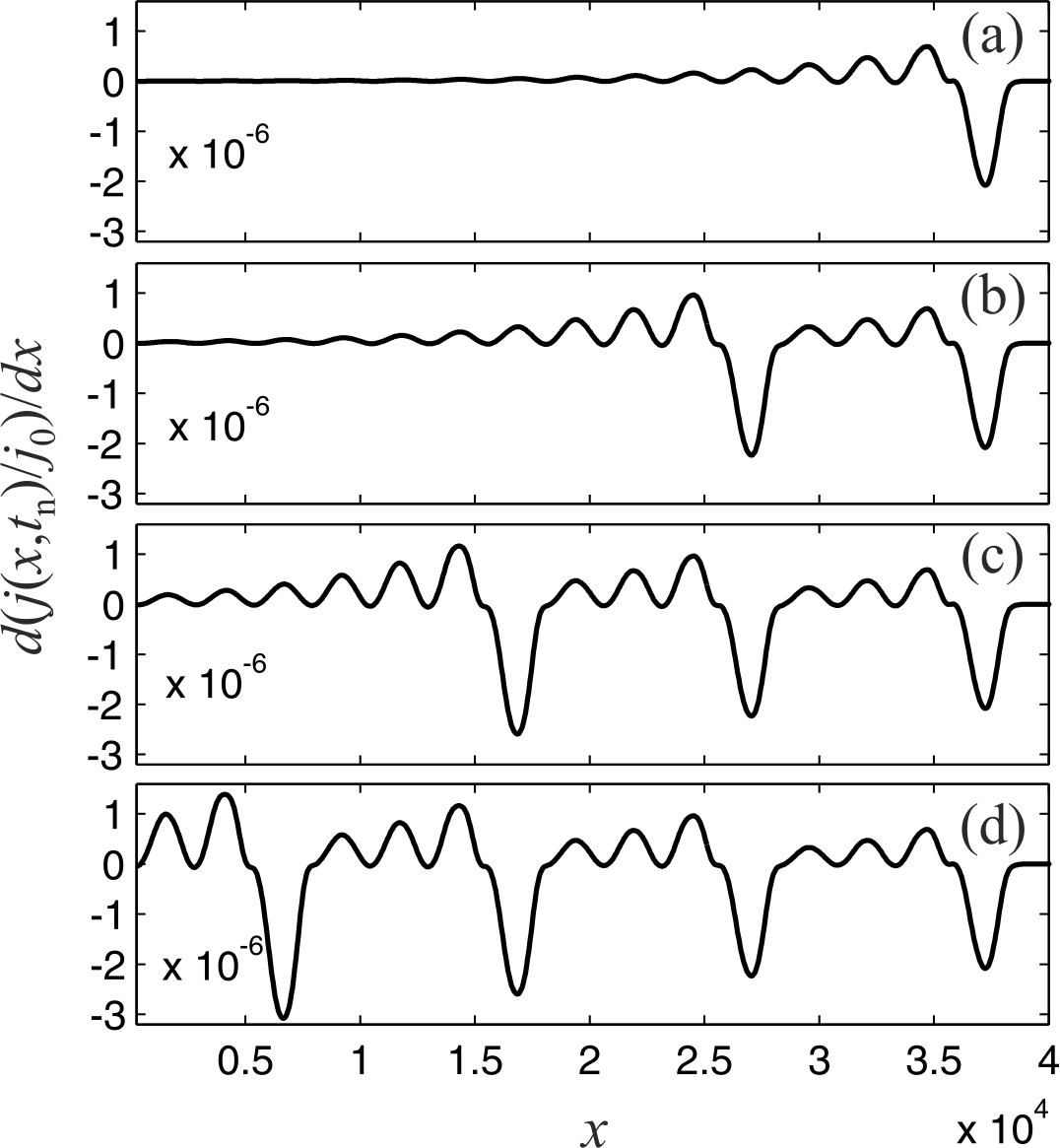}\\
  \caption{ Time dependence of the derivatives with respect to the time on the sequences of photocurrent density pulses shown in Fig.\ref{FIG:Fig7} at the exit point $x_3=2d$  from the heterostructure. Time in atomic units  ($t$) a.u. $ = 2.419 \times 10^{ - 17} $ s.}\label{FIG:Fig8}
\end{figure}

\begin{figure}[h]
\includegraphics[width=6.5 cm]{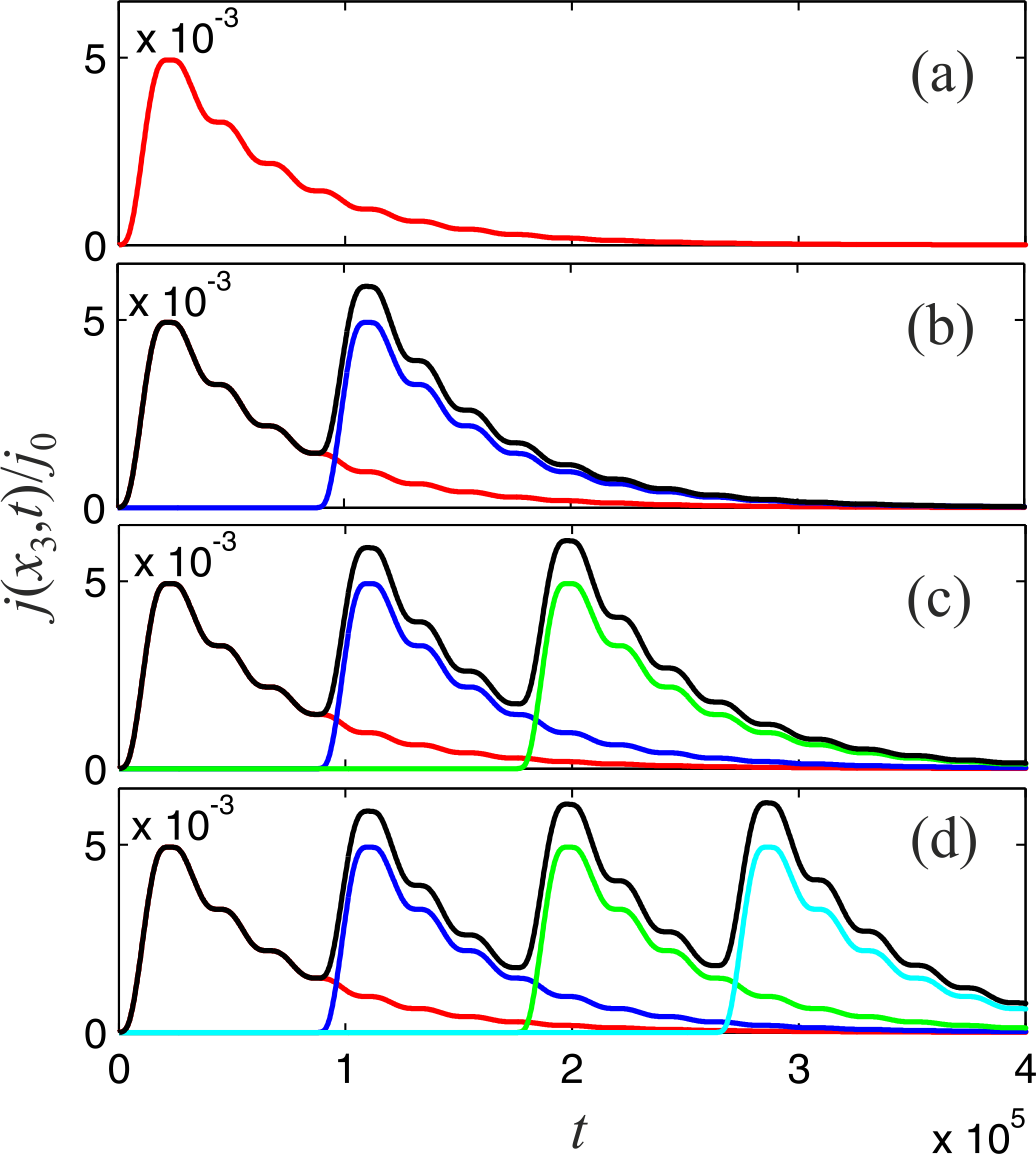}\\
  \caption{(Color online) The coordinate dependence of the current density pulses from a photocathode with a surface heterostructure calculated at the instant of time $t = 15T_{12} $, caused by the action of (a) one, and sequences (b) two, (c) three, (d) four identical light pump pulses with a duration $t_0=T_{12}$  shifted in time by  $\delta t=4T_{12}$. Coordinate  $x$ in angstroms \AA.}\label{FIG:Fig9}
\end{figure}

\begin{figure}[h]
\includegraphics[width=6.5 cm]{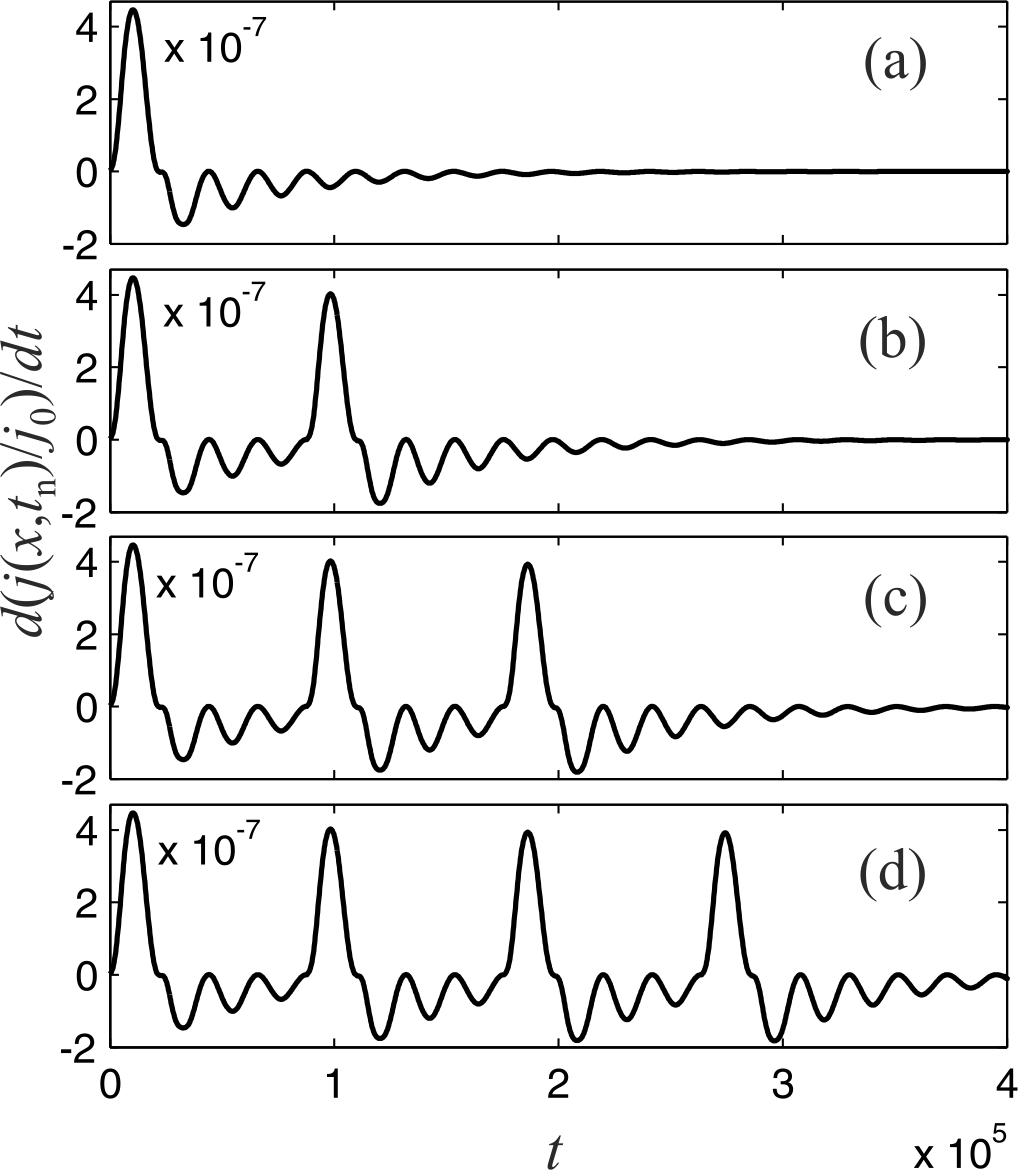}\\
  \caption{ Coordinate dependence of the derivatives with respect to the coordinate on the sequences of current density pulses from the photocathode with a surface heterostructure, shown in Fig.\ref{FIG:Fig9}, calculated at the instant of time  $t = 15T_{12} $. Coordinate $x$ in angstroms \AA.}\label{FIG:Fig10}
\end{figure}

Figures (Fig.\ref{FIG:Fig7} and (Fig.\ref{FIG:Fig9}) demonstrate such a coherent prolongation of generation with amplification of the photocurrent density waves by a sequence of four ($N=4$) identical pump light pulses with a duration   $t_0=T_{12}$ shifted in time by  $\delta t=4T_{12}$. Figures (Fig.\ref{FIG:Fig8}) and (Fig.\ref{FIG:Fig10}) demonstrate the manifestation of the discussed spatio-temporal oscillations with the difference wave harmonics of the period $T_{12}$  and wavelength $\lambda_{12}$  through the behavior of the corresponding first derivatives of the photocurrent density with respect to time and coordinate.

\section{CONCLUSION}

In our work, we developed the theory of nonstationary photoemission, based on the method of the density matrix of one-electron states, applicable for fast photoemission processes at times shorter than the times of inelastic relaxation of electrons, in particular, in thin-film photocathodes. The parameters of the theory are one-electron wave functions, energy spectrum and characteristics of smearing of electronic states due to rather weak inelastic scattering. The theory makes it possible to calculate the coordinate-time dependence of the pulses of the charge and current densities taking into account weak relaxation processes inside the photocathode and the presence of a surface heterostructure. In the limit of stationary photoemission under the action of monochromatic light, the theory occupies an intermediate position between the three-step Spicer model and the one-step quantum model.

The calculations of the alternating photoemission current were carried out for a simplified scheme of a planar photocathode with a surface heterostructure in the form of a double quantum well, which serves as a filter for photoelectrons. For a photoelectron wave packet with an optimal energy width, it can provide spatio-temporal wave-like modulation of charge and current densities with a frequency and wavelength that correspond to the difference frequency of the transition between the resonance levels of quasi-stationary states of the surface three-barrier heterostructure. A wave packet can be formed using electric and magnetic fields of the appropriate configuration, by extracting electrons from the photocurrent, the energies of which belong to a band wider than the energy distance between the levels of a certain doublet, but narrower than the distance to neighboring doublets.

For efficient generation of the difference harmonic component of the alternating photocurrent, it is required that the duration of the pump pulses should be shorter than the relaxation time, and the intervals between them should be shorter than the lifetime of quasi-stationary states, which can be large in thin quantum-well films. The characteristics of the photocurrent pulses strongly depend on the parameters of the heterostructure. 
For layer thicknesses of the three-barrier heterostructure of 
1 - 10 nm and barrier heights of 0.5 - 2.5 eV, 
the lifetimes of quasi-stationary states of 10$^{-1}$-10$^2$ and the generated difference frequencies for them of $10^{11}-10^{14}$ Hz can be provided. 
It is possible to change the lifetimes $\tau _{R1} ,\tau _{R2} $
  and difference frequencies $\nu_{12}$  of doublet quasi-stationary states by varying the parameters of the surface heterostructure, which changes the shape of the
curves of the photocurrent versus time; the analysis of these curves can also provide information on the values of the relaxation times $\tau _p  = \hbar /\gamma _p$
 of excited electrons in the photocathode. With the formation of a positive feedback between the pulses of the photocurrent and the light source with the transition of the system to the self-oscillation mode, based on the described effect, it is possible to create a current generator in the terahertz frequency range.

\appendix*
\section{}

We replace the sums over  $p_1$ in expressions \eqref{eq:math:19} and \eqref{eq:math:21} by integrals over $\xi_{p_1}$  and write $f(\omega ,p,p',p_1)$
  from (20) in the form
   \begin{equation}\label{eq:math:A1}
f(z) = \frac{1}
{{z - z_1 }} - \frac{1}
{{z - z_2 }} = \frac{{z_1  - z_2 }}
{{(z - z_1 )(z - z_2 )}}
\end{equation}
where  $z = \xi _{p_1 } ,\;\;z_1  = \xi _p  - \hbar \omega  + i\gamma _p ,\;\;z_2  = \xi _{p'}  - \hbar \omega  - i\gamma _{p'} ,\;\;z_1  - z_2  = \xi _p  - \xi _{p'}  + i\gamma _{pp'} $. The function $f(z)$   has a pole $z_1$  in the upper half-plane and a pole $z_2$  in the lower 
half-plane of the complex variable  $z$.

It can be assumed that the width $\Delta \varepsilon _{p_1 } $
  of the lower band of unexcited states is large in comparison with the distance between the levels of the resonance doublet and with the width of the recorded energy band  $E_{R2}  - E_{R1}  < E_{\max }  - E_{\min }  \ll \Delta \varepsilon _{p_1 } $; therefore, we extend the rapidly converging integrals over $\xi _{p_1 }  = z$
  to the entire real axis  $ - R < z <  + R$,  $R \to \infty $. We close the corresponding integral contours in the upper or in the lower half-plane with semicircles of large radius  $R$, the contribution of which tends to zero at   $R \to \infty $, and we find the residues at the corresponding poles. As a result, we find that the first sum in \eqref{eq:math:19} and the sum \eqref{eq:math:21} are approximated by the integral
\begin{equation}\label{eq:math:A2}
\sum\limits_{p_1 } {D_{p_1 } } f(\omega ,p,p',p_1 ) \approx D\int\limits_{ - \infty }^{ + \infty } {f(z)} dz = 2\pi iD
\end{equation}
and the second sum in \eqref{eq:math:19} is approximated by the integrals
\begin{widetext}
   \begin{equation}\label{eq:math:A3}
\begin{gathered}
  \sum\limits_{p_1 } {D_{p_1 } } f(\omega ,p,p',p_1 )\left. {\left[ {e^{i\left( {\hbar \omega  - (\xi _{p'}  - \xi _{p_1 } )} \right){t \mathord{\left/
 {\vphantom {t \hbar }} \right.
 \kern-\nulldelimiterspace} \hbar } - \gamma _{p'} {t \mathord{\left/
 {\vphantom {t \hbar }} \right.
 \kern-\nulldelimiterspace} \hbar }}  + e^{ - i\left( {\hbar \omega  - (\xi _p  - \xi _{p_1 } )} \right){t \mathord{\left/
 {\vphantom {t \hbar }} \right.
 \kern-\nulldelimiterspace} \hbar } - \gamma _p {t \mathord{\left/
 {\vphantom {t {\hbar }}} \right.
 \kern-\nulldelimiterspace} {\hbar }}} } \right]} \right\} \approx  \hfill \\
  \quad  \approx D\left[ {\int\limits_{ - \infty }^{ + \infty } {f(z)} e^{i(z - z_2 ){t \mathord{\left/
 {\vphantom {t {\hbar }}} \right.
 \kern-\nulldelimiterspace} {\hbar }}} dz + \int\limits_{ - \infty }^{ + \infty } {f(z)} e^{ - i(z - z_1 ){t \mathord{\left/
 {\vphantom {t {\hbar }}} \right.
 \kern-\nulldelimiterspace} {\hbar }}} dz} \right] = 4\pi iDe^{i(z_1  - z_2 ){t \mathord{\left/
 {\vphantom {t {\hbar }}} \right.
 \kern-\nulldelimiterspace} {\hbar }}} . \hfill \\ 
\end{gathered} 
\end{equation}
\end{widetext}
After substituting the right-hand sides \eqref{eq:math:A2} and \eqref{eq:math:A3} in \eqref{eq:math:19} and \eqref{eq:math:21}, we arrive at simple expressions \eqref{eq:math:23} and \eqref{eq:math:24}.

\end{document}